\documentclass[aps, prb, reprint, superscriptaddress, longbibliography, amsmath, amssymb]{revtex4-2}

\usepackage{graphicx}
\usepackage{bm}
\usepackage{hyperref}
\usepackage{graphicx}
\usepackage{siunitx}

\newcommand{\Bc}{\ensuremath{B_1}}
\newcommand{\gammah}{\ensuremath{\gamma}}

\begin{document}

\title{Magnomechanics in suspended magnetic beams}

\author{Kalle S.~U.~Kansanen}
\email{kalle.s.u.kansanen@jyu.fi}
\affiliation{Department of Physics and Nanoscience Center, University of Jyv\"akyl\"a, P.O. Box 35 (YFL), FI-40014 University of Jyv\"askyl\"a, Finland}

\author{Camillo Tassi}
\affiliation{Department of Physics and Nanoscience Center, University of Jyv\"akyl\"a, P.O. Box 35 (YFL), FI-40014 University of Jyv\"askyl\"a, Finland}

\author{Harshad Mishra}
\affiliation{Department of Applied Physics, Aalto University, P.O. Box 15100, 00076 Aalto, Finland}

\author{Mika A.~Sillanp\"a\"a}
\affiliation{Department of Applied Physics, Aalto University, P.O. Box 15100, 00076 Aalto, Finland}

\author{Tero T.~Heikkil\"a}
\affiliation{Department of Physics and Nanoscience Center, University of Jyv\"akyl\"a, P.O. Box 35 (YFL), FI-40014 University of Jyv\"askyl\"a, Finland}

\date{December 7, 2021}

\begin{abstract}
Cavity optomechanical systems have become a popular playground for studies of controllable nonlinear interactions between light and motion. Owing to the large speed of light, realizing cavity optomechanics in the microwave frequency range requires cavities up to several mm in size, hence making it hard to embed several of them on the same chip. An alternative scheme with much smaller footprint is provided by magnomechanics, where the electromagnetic cavity is replaced by a magnet undergoing ferromagnetic resonance, and the optomechanical coupling originates from 
magnetic shape anisotropy. Here, we consider the magnomechanical interaction occurring in a suspended magnetic beam -- a scheme in which both magnetic and mechanical modes physically overlap and can also be driven individually. We show that a sizable interaction can be produced if the beam has some initial static deformation, as is often the case due to unequal strains in the constituent materials. We also show how the magnetism affects the magnetomotive detection of the vibrations, and how the magnomechanics interaction can be used in microwave signal amplification. Finally, we discuss experimental progress towards realizing the scheme.
\end{abstract}

\maketitle

\section{Introduction}

In cavity optomechanical devices, the radiation pressure force mediates an interaction between mechanical modes and photons~\cite{aspelmeyer2014cavity}. This has led to several developments, both fundamental and applied: ground state cooling \cite{Teufel2011b,AspelmeyerCool11} and entanglement \cite{Entanglement,Groblacher} of mechanical modes, 
quantum information storage in the mechanical excitation and interface, e.g.~between superconducting qubits and flying optical photons \cite{Painter2020scqb}, sensitive measurements with precision limits given by quantum mechanics; classical signal processing with tunable nonlinearities \cite{Teufel2016fConv,CasparAmp,Verhagen2016isol,Kiina2016NonRes,Floquet2020}.

Several implementations of cavity optomechanics have been considered: mirrors on cantilevers and beams ~\cite{Karrai2004,Aspelmeyer2006cool}; membranes in cavities \cite{thompson2008strong,Schliesser2018FB}; atomic clouds \cite{Stamper2010PRL};  beams and plate capacitors in microwave resonators \cite{Lehnert2008Nph,Schwab2010,Teufel2011b,Teufel2016fConv,CasparAmp,Floquet2020}; photonic crystals patterned into beams~\cite{AspelmeyerCool11}.

Recently, the coupling between magnons and phonons has been considered to obtain an interaction similar to that in cavity optomechanics, but with the role of photons now played by magnons ~\cite{zhang2016cavity,gonzalez2020theory,gonzalez2020quantum,potts2021dynamical}.
The interaction between magnons and phonons is
mediated by the combination of the magnetic shape anisotropy and the magnetoelastic effect 
which make the frequency of the magnon, i.e.,~the ferromagnetic resonance (FMR), dependent on the strain. One interesting feature is that the speed of spin waves is substantially lower than that of electromagnetic waves, which can offer a much denser integration of similar functionalities. Moreover, enhanced tunability and richer interaction suggest additional possibilities for devices for fundamental studies and applications.

\begin{figure}
\centering
{\includegraphics{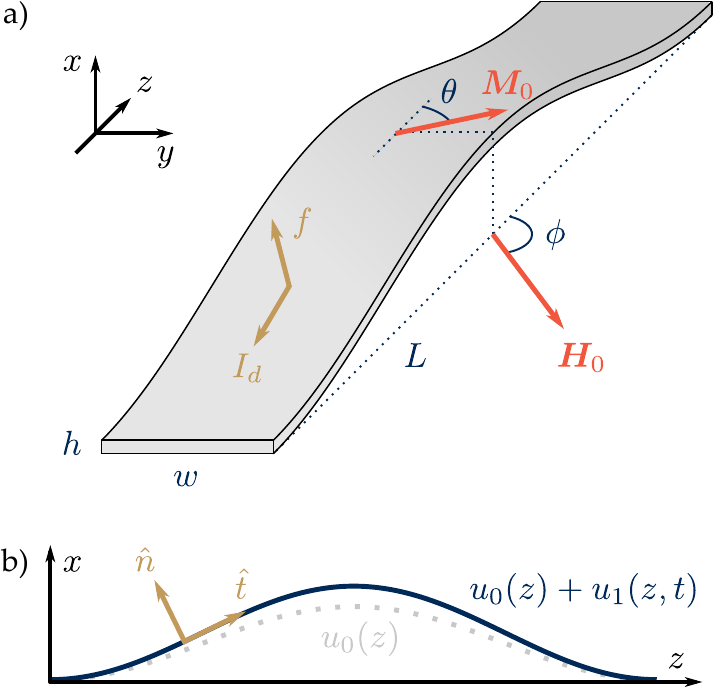}} 
\caption{ Schematic drawings of a doubly clamped magnetic beam with a static deformation. 
a) A three dimensional perspective drawing of a static suspended beam with several useful definitions (see main text).
b) A simplified view of the deformation of the beam axis from $y$ direction.
The gray dotted line represents the static deformation $u_0(z)$ and the blue solid line the total displacement $u = u_0 + u_1$ that includes the time-dependent vibrations $u_1$.
Here, $\hat{t}$ and $\hat{n}$ are tangent and normal vectors of the beam, corresponding to the local direction of the current $I_d$ and the force $f$, respectively.}
\label{fig:beamsketch}
\end{figure}

We propose and theoretically describe a scheme for magnomechanics that is a direct analog of a basic microwave optomechanical system where a suspended conducting beam is capacitively coupled to a microwave resonator.
We consider a suspended ferromagnetic beam that is subjected to an external in-plane magnetic field.
This generates the FMR or magnon mode which is affected by the vibrations of the beam.
However, as shown below, a simple doubly clamped beam does not give rise to the desired magnomechanical coupling.
Since the FMR frequency changes as much to upward as to downward displacements of the beam, the analog of the radiation pressure force vanishes.
Such symmetry may be broken by a static deformation as in Fig.~\ref{fig:beamsketch}a.

Following Fig.~\ref{fig:beamsketch}a, we consider a ferromagnetic beam, magnetized as a single domain and having a rectangular cross section with width $w$ and height $h$.
We assume that the coordinate system may be oriented so that the beam displacement is in the $x$ direction and the beam axis is in the $z$ direction. 
The clamps fix the ends of the beam to be in the $xy$ plane parallel to one another with distance $L$ in the $z$ direction.
The external and static magnetic field $\bm{H}_0$ is assumed to lay in the $yz$ plane.
The external field $\bm{H}_0$ fixes the static magnetization $\bm{M}_0$ around which magnons are generated.

The setup with a doubly clamped beam allows for separate driving of mechanical and ferromagnetic resonance modes.
The measurement scheme is conceptually similar to microwave optomechanics.
The mechanical mode can be driven by driving a current $I_d$ through the beam -- assuming that the beam is conducting -- which generates a Lorentz force $f$ on the beam due to the external field $\bm{H}_0$.
Similarly, the magnon mode can be driven by a time-dependent magnetic field $\bm{h}$, for instance, generated by a microwave antenna.
These schemes are commensurate with the typical reflection and transmittance measurements performed with vector network analyzers.
Moreover, the separate control over mechanics and magnons opens a perspective to signal transduction.

In contrast to recent experiments~\cite{zhang2016cavity, potts2021dynamical}, this setup does not require a cavity to hybridize with the magnon mode, lessening the physical size of the magnomechanical system.
Also, it does not require any magnetic field gradients as in a few recent theoretical works~\cite{gonzalez2020theory,gonzalez2020quantum}; the fields $\bm{H}_0$ and $\bm{h}$ are here considered to be homogeneous in space.

Magnomechanics is in many ways similar to optomechanics but there are several points of distinction.
In the setup of Fig.~\ref{fig:beamsketch}, it is possible to modulate the optomechanical coupling by tuning either the strength or the direction of the external magnetic field $\bm{H}_0$.
This in turn changes the response of the system to driving.
We also focus on a feature unique to magnetic systems: magnetic hysteresis and its effect on the magnomechanical coupling.

The  layout  of  the  paper  is  as  follows. 
In  Sec.~II we derive the magnomechanical Hamiltonian for a statically deformed beam. 
In Sec.~III we describe the driving of the mechanical and ferromagnetic resonances
which is used in Sec.~IV to describe possible experiments for the system.
Section V discusses the first steps towards implementing a doubly clamped magnetic beam as a platform for magnomechanics.

\section{Magnomechanical Hamiltonian}

To motivate the upcoming discussion, let us discuss the optomechanical Hamiltonian in detail.
The simplest derivation of optomechanics follows from the assumption of a cavity, described by a bosonic mode $\hat c$, whose eigenfrequency $\omega_c$ depends on the position $\hat{x}_b$ of some mechanical mode $\hat b$ (e.g. a mirror connected to a spring).
The typical optomechanical Hamiltonian is then obtained by an expansion of the cavity frequency to first order in the position $\hat{x}_b$
\begin{eqnarray}
 \hat{H}_{\text{optm}} &&=
  \hbar \omega_{\text{m}} \hat{b}^\dagger \hat{b} + \hbar \omega_{c}(\hat{x}_b) \hat{c}^\dagger \hat{c}\nonumber\\
 &&\simeq \hbar \omega_{m} \hat{b}^\dagger \hat{b}
 + \hbar \omega_{c} \hat{c}^\dagger \hat{c}
 + \hbar g_0 \hat{c}^\dagger \hat{c} (\hat{b}^\dagger+\hat{b}),
 \label{eq:optom_ham}
\end{eqnarray}
where $\omega_m$ is the eigenfrequency of the mechanical mode.
The constant $g_0$ is called the optomechanical coupling and the corresponding term describes the radiation pressure force.

A similar mechanism can be found in ferromagnetic systems.
Then, the cavity is replaced by a ferromagnetic resonance (FMR) and $\hat c$ corresponds to a magnon mode.
There are at least two distinct physical mechanisms that allow the FMR frequency to depend on a mechanical mode: 
On one hand, it is well known that the shape of the ferromagnet affects its FMR frequency~\cite{kittel1948fmr}.
On the other hand, the magnetoelasticity directly changes the ferromagnetic dynamics, providing an extraneous anisotropy field~\cite{kittel1949physical}. 
Therefore, the relevant mechanical mode $\hat b$ is the vibration of the ferromagnet itself.

In this section, we start by describing the mechanical and ferromagnetic dynamics of a deformed beam from which we derive the magnomechanical coupling that corresponds to $g_0$ in Eq.~\eqref{eq:optom_ham}.

\subsection{Mechanical dynamics}

We describe the deformations of the beam in Fig.~\ref{fig:beamsketch} within the Euler-Bernoulli beam theory that is typically used to describe beam dynamics, even in the nanometer scale~\cite{heikkila2013physics}.
In order to describe the large static deformation ($h < \max u_0(z)$ in Fig.~\ref{fig:beamsketch}b) and vibrations around it, we include in the Euler-Bernoulli equation nonlinear terms that are typically neglected~~\cite{landau1986theory, reddy2013generalized}. 
We assume that the beam material is homogeneous and isotropic.
Then, vibrations in different directions do not generally couple, and the following analysis can be done separately for vibrations in the $x$ and $y$ directions.
However, it turns out that the static deformation is needed for a finite magnomechanical coupling.
In the following, we thus neglect the beam dynamics in the $y$ direction.
We also neglect shear modes, that is, modes where the displacement in the $x$ direction depends on $y$, assuming that the width $w$ is appreciably smaller than the length $L$.
These assumptions are consistent with the actuation scheme of Lorentz force in Fig.~\ref{fig:beamsketch}.
Without any external force, the beam displacement $u(z,t)$ in the $x$ direction obeys the equation~\cite{reddy2013generalized,nayfeh2008exact}
\begin{equation}
 \rho A \frac{\partial^2 u}{\partial t^2} + E  I_x \frac{\partial^4 u}{\partial z^4} 
 =\left[T_0 + \frac{E A}{2 L} \int_0^L dz \left( \frac{\partial u}{\partial z} \right)^2  \right] \frac{\partial^2 u}{\partial z^2},
 \label{eq:non-linear_EB}
\end{equation}
where $T_0$ is the initial tension, $\rho$ is the mass density, $E$ the Young modulus, $A=w h$ the area of the beam with a rectangular cross section, and $I_x = w h^3/12$ the corresponding bending modulus. 
Here, the terms on the right provide the nonlinear corrections related to tension whereas the usual Euler-Bernoulli kinetic and stress terms are on the left.

In order to describe the initially buckled beam and the flexural vibrations around it, we choose an approach similar to Ref.~\onlinecite{nayfeh2008exact} and discuss it briefly here.
Further details can be found in Appendix~\ref{sec:mechanical_eigenmodes}.

A static deformation of the beam can be caused by a compressive stress that is, for instance, due to an external force on the clamping mechanism or due to the fabrication method of the beam.
Here, we assume that this compression may be described by a negative tension $T_0 < 0$ in Eq.~\eqref{eq:non-linear_EB}.
We then divide the total mechanical deformation $u$ into a static and dynamical part ($u_0$ and $u_1$, respectively) as in Fig.~\ref{fig:beamsketch}b and  assume that the dynamical deformation is small, that is,
\begin{equation}
\label{eq:beam_boundary_cond}
    u(z,t) = u_0(z) + u_1(z,t),\qquad u_1 \ll u_0.
\end{equation}
This allows for an expansion of the non-linear Euler-Bernoulli equation~\eqref{eq:non-linear_EB} in the powers of $u_1$.
Together with the boundary conditions of a doubly clamped beam for both the static and dynamic deformations ($\partial_{z}$ denotes $z$ derivative)
\begin{equation}
 u_{0/1}(0)= u_{0/1}(L)= \partial_{z}u_{0/1}(0) = \partial_{z}u_{0/1}(L) =  0,
\label{eq:non-linear_boundary_cond}
\end{equation}
the mechanical system is now fully described.

The static deformation $u_0$ is solved from the zeroth order equation in $u_1$.
We find a particularly simple expression
\begin{equation}
    u_0(z) = u_m\frac{1 - \cos(2\pi z/L)}{2},
    \label{eq:staticbeam}
\end{equation}
where the parameter $u_m$ gives the displacement at the mid-point of the beam as $u_0(L/2) = u_m$.
This is also the point of the largest displacement.
There is a one-to-one correspondence between $u_m$ and the negative tension $T_0$, which is why they can be used interchangeably.

From the first order equation for the dynamic deformation $u_1$ -- neglecting the terms that are of second order in $u_1$ -- we can then find the flexural eigenmodes $\chi_n$ and the corresponding eigenfrequencies $\omega_n$.
These do not admit to simple expressions but a numerical solution is readily available~\cite{numericsnote}.
We find that the dimensionless eigenfrequencies $\bar{\omega}_n = \frac{L^2}{h}\sqrt{\frac{12\rho}{E}}\omega_n$ are determined by the ratio of the largest static deformation to the height of the beam $u_m/h$. 
Especially, the eigenfrequencies of symmetric eigenmodes ($n= 1, 3 \dots$) increase with $u_m/h$ while the eigenfrequencies of antisymmetric modes ($n= 2, 4 \dots$) are constant (see Fig.~\ref{fig:mech_eigenfr} in App.~\ref{sec:mechanical_eigenmodes}).

With the assumption of small dynamical deformation, the general solution can be expressed as a superposition
\begin{equation}
    \label{eq:u1}
    u_1(z,t) =
    \sum_n x_n(t) \chi_n(z)
\end{equation}
with the dynamical amplitudes $x_n$ and eigenmodes $\chi_n$ normalized as $\int_0^L \chi_n \chi_m dz = L\delta_{ij}$.
This in turn gives the mechanical energy
\begin{equation} \label{eq:mechanicalH}
    H_{mc} = \sum_n \left[ \frac{p_n^2}{2 m} + \frac{1}{2} \omega_n^2 x_n^2 \right],
\end{equation}
where the mass is given by $m=\rho L A$ and the momentum by $ p_n = m \dot{x}_n$.
Then, we can quantize the harmonic oscillators as usual:
\begin{subequations}
\begin{eqnarray}
   &&\hat {x}_n=x^{\rm ZPM}_n
   (\hat{b}_n^{\dagger }+\hat{b}_n),\\
   &&\hat{b}_n^{\dagger}=(x^{\rm ZPM}_n)^{-1}\left({\hat{x}_n}-{\frac{i}{ m \omega_n }}{\hat {p}_n}\right),
\end{eqnarray}
\end{subequations}
where $x^{\rm ZPM}_n={\sqrt {\frac {\hbar }{2 m \omega_n }}}$ is the
zero-point motion amplitude and $\hat{b}_n^\dagger,\,\hat{b}_n$ are boson operators for each mode $n$.

\subsection{Ferromagnetic resonance}

We describe the magnetization $\bm{M}$ and the magnon mode of the beam in Fig.~\ref{fig:beamsketch} assuming that it is magnetized as a single domain, similar to the Stoner-Wohlfarth model~\cite{stoner1948mechanism}.
The magnetization is driven by an effective field $\bm{H}_\mathrm{eff}$
that takes into account the
external magnetic field $\bm{H}$, the demagnetizing field
$\bm{H}_{\mathrm{dm}}$, the magnetoelastic field
$\bm{H}_{\mathrm{me}}$, and intrinsic anisotropy field $\bm{H}_\mathrm{an}$~\cite{gilbert2004phenomenological}. 
We consider the magnetization almost uniform, adiabatically following the beam geometry, and that the magnet is in the Landau-Lifshitz-Gilbert regime where $\lvert \bm{M}\rvert=M_S$ with $M_S$ being the saturation magnetization. 
For the external magnetic field $\bm{H}$, we  consider a strong static component $\bm{H}_0$ and a perturbation $\bm{h}$ oscillating at a frequency close to the ferromagnetic resonance (FMR) frequency $\omega_K$, although the latter plays a role only in the following section.  
Equivalently, the magnetization dynamics can be obtained from the magnetic free energy density within the beam
\begin{subequations}
\label{eq:megn_energy_eff_field}
\begin{align}
    \mathcal{F} &= -\mu_0 \bm{H}\cdot \bm{M} + \mathcal{F}_\mathrm{dm} + \mathcal{F}_\mathrm{me} + \mathcal{F}_\mathrm{an} \\
     \mu_0 \bm{H}_{\text{eff}} &=-\frac{\delta}{\delta \bm{M}} \int dx^3\, \mathcal{F}[\bm{M}].
\end{align}
\end{subequations}
To obtain the total magnetic energy, the free energy density $\mathcal{F}$ must be integrated over the beam.
Next, we describe the terms in the free energy density in detail for the deformed beam.
We first discuss the results arising from the assumption that there are no vibrations in $y$ direction (meaning that the displacement $u$ is only in $x$ direction and it is a function of $z$).
Then, we return to assess this assumption.

We calculate the demagnetizing field $\bm{H}_\mathrm{dm}$ in the thin plate limit, 
$h \ll w \lesssim L$, which allows us to neglect the size-dependent terms in the demagnetizing field.
It is obtained by requiring the continuity of the $H$-field components parallel to the beam, that is, $H_y$ and the tangent field $H_t=[H_x \partial_z u(z) + H_z]/\sqrt{1+(\partial_z u(z))^2}$, and the continuity of the $B$-field component normal to the beam, that is $B_n=[-B_x + B_z \partial_z u(z)]/\sqrt{1+(\partial_z u(z))^2}$ (see Fig.~\ref{fig:beamsketch}), where $\bm{B}=\mu_0\, \bm{H}^{\text{out}}$ outside the beam and $\bm{B}=\mu_0\, (\bm{H}^{\text{in}}+\bm{M})$ inside. 
The demagnetizing field $\bm{H}_\mathrm{dm}$ is within the beam and, thus, we set $\bm{H}^\mathrm{out} = \bm{H}$ and $\bm{H}^\mathrm{in} = \bm{H} + \bm{H}_\mathrm{dm}$.
Then, up to the second order in $\partial_z u$, we find its components
\begin{subequations}
\begin{eqnarray}
    &&H_x^{\text{dm}}=-M_x [1-(\partial_z u(z))^2] + M_z \partial_z u(z),\\
    &&H_y^{\text{dm}}=0,\\
    &&H_z^{\text{dm}}=M_x \partial_z u(z) - M_z (\partial_z u(z))^2
\end{eqnarray}
\end{subequations}
from the continuity conditions.
The corresponding free energy density is given by
\begin{equation}
\label{eq:demagnetizing_field_energy}
    \frac{\mathcal{F}_{\text{dm}}}{\mu_0} =
 \frac{M_x^2}{2} - M_x M_z \partial_z u(z)
 +\frac{1}{2} \left(M_z^2-M_x^2\right) (\partial_z u(z))^2.
\end{equation}
It describes a hard axis parallel to $\hat{n}$, the
(position-dependent) surface normal to the beam.   
Since we have to integrate the free energy density due to the
demagnetizing field over the beam volume, the  component proportional to
$\partial_z u(z)$ disappears in the case of a doubly clamped beam due to
the boundary conditions $u(0)=u(L)=0$~\cite{cantilevernote}.
Moreover, for a small deformation the averaged surface
normal aligns with the $x$ direction, which is why $x$ axis is a hard axis for a uniform magnetization. 
This means that for an in-plane magnetic field considered below, the magnetization also lies in the $yz$ plane. 
On the other hand, deformations provide a position dependent correction to this hard axis. 
These corrections are sensitive to vibrations and thus provide one contribution to the magnomechanical coupling.

In general, the magnetoelastic free energy density is given by~\cite{kittel1949physical,sander1999correlation}
\begin{align}
    \mathcal{F}_\mathrm{me} =\, &\frac{B_1}{M_S^2}(M_x^2 \epsilon_{xx} + M_y^2 \epsilon_{yy} + M_z^2 \epsilon_{zz}) \nonumber\\
    &+\frac{2B_2}{M_S^2} (M_x M_y \epsilon_{xy} + M_x M_z \epsilon_{xz} + M_y M_z \epsilon_{yz}),
\end{align}
where $B_1$ and $B_2$ are the magnetoelastic coupling constants, and the strains $\epsilon_{ij}$ are given in the non-linear Euler-Bernoulli theory by~\cite{landau1986theory, strainnote}
\begin{align}
    \epsilon_{ij} = \frac{1}{2}\left( \frac{\partial u_i}{\partial x_j} + \frac{\partial u_j}{\partial x_i} + \sum_k\frac{\partial u_k}{\partial x_i}\frac{\partial u_k}{\partial x_j} \right).
\end{align}
Here, $u_i$ are the components of the general displacement vector at a given point. 
The last term in $\epsilon_{ij}$ is neglected in the linear elasticity theory but is here of importance.

Let us then consider, for convenience, the average of $\mathcal{F}_\mathrm{me}$ over the beam volume and denote it by $\bar{\mathcal{F}}_\mathrm{me}$.
Since $u_x \equiv u$ and $u_y = u_z = 0$, only the derivatives $\partial_z u$ and $\partial_y u$ are finite.
Moreover, setting $\partial_y u = 0$ for now, only $\epsilon_{zz}$ and $\epsilon_{xz}$ are finite.
However, in the process of averaging over the beam volume, we encounter the same situation as in the demagnetizing energy: first order terms $\partial_z u$ average out due to boundary conditions.
Thus, we find
\begin{align}
    \bar{\mathcal{F}}_\mathrm{me} = \int \frac{dx^3}{L A} \mathcal{F}_\mathrm{me}
    = \frac{B_1}{M_S^2} M_z^2 \bar{\epsilon}_{zz}
\end{align}
with the definition of averaged strain
\begin{align}
    \bar{\epsilon}_{zz} = \frac{1}{L}\int d z \epsilon_{zz} = \frac{1}{2L}\int d z (\partial_z u)^2.
    \label{eq:averagestrain}
\end{align}
Note that the total demagnetizing energy is also proportional to $\bar{\epsilon}_{zz}$ due to the last term in Eq.~\eqref{eq:demagnetizing_field_energy}.
Moreover, by using the expansion~\eqref{eq:beam_boundary_cond} of the displacement $u$ into a static deformation and dynamical vibrations, the average strain can be expanded to first order by $\bar{\epsilon}_{zz} \simeq \bar{\epsilon}_{zz}^{(0)} + \bar{\epsilon}_{zz}^{(1)}$ with
\begin{subequations}\label{eq:strainexpansion}
\begin{eqnarray}
   &&\bar{\epsilon}_{zz}^{(0)}=\frac{1}{2 L}\, \int_0^L dz\, (\partial_z u_0)^2,\\
     &&\bar{\epsilon}_{zz}^{(1)}=\frac{1}{L}\, \int_0^L dz\, (\partial_z u_0) (\partial_z u_1). 
\end{eqnarray}
\end{subequations}
Using the solution of the static deformation in Eq.~\eqref{eq:staticbeam}, we obtain $\bar{\epsilon}_{zz}^{(0)} = \frac{\pi^2}{4} (u_m/L)^2$.
Here, we expect that $u_m \ll L$ meaning that $\bar{\epsilon}_{zz}^{(0)} \ll 1$.

Finally, we assume that the magnetoelastic and demagnetizing fields dominate over the intrinsic (crystal) anisotropies, that is, the total free energy of the intrinsic anisotropies is small compared to the described free energies and may be neglected.
At least, such an assumption is valid in the case of a polycrystalline magnet. 
This assumption may be lifted straightforwardly in favor of a specific crystal anisotropy model; however, difficulties may arise when trying to take into account the possible direct effect of elasticity on the intrinsic anisotropies.

The average magnetic free energy density, with the assumption of the displacements only in the $x$ direction, is 
\begin{align}\label{eq:averagemagneticF}
    \bar{\mathcal{F}} = -&\mu_0 \bm{H}\cdot \bm{M}
     + \frac{\Bc}{M_S^2} M_z^2 \bar{\epsilon}_{zz} \nonumber\\
      +&\mu_0 M_x^2/2 +\mu_0 \left(M_z^2-M_x^2\right) \bar{\epsilon}_{zz}.
\end{align}
The addition of beam vibrations in the $y$ direction only provides an additional term to $\bar{\epsilon}_{zz}$ that is proportional to $(\partial_z u_y)^2$.
However, such term does not provide the correct form of optomechanical coupling as in Eq.~\eqref{eq:optom_ham}: it is second order in the position operator.
In contrast, the contribution of vibrations in the $x$ direction around the static deformation $u_0$ is linear in position because of the expansion in Eq.~\eqref{eq:strainexpansion} and $\bar{\epsilon}_{zz}^{(1)}$.
Finally, we note that if $\partial_y u \neq 0$, both $\epsilon_{yz}$ and $\epsilon_{yy}$ are also finite.
Without a static deformation in the $y$ direction, $\bar{\epsilon}_{yy}$ does not provide magnomechanical coupling but $\bar{\epsilon}_{yz}$ will via the nonlinear term $(\partial_y u)(\partial_z u_0)$.
Similar terms would appear from the demagnetizing field.
Typically, such shear terms are neglected in optomechanics: this is justified if the beam width is much smaller than its length, or that only modes without shear are actuated.
We follow the same assumption in this work.

We treat both cases of the metallic and insulator ferromagnets 
(e.g.~a two-layer beam with a ferromagnetic insulating layer and a metal layer). 
To do so, we can observe that the current induced by a static field $\bm{H}_0$ and the flexural dynamics gives rise to a negligible screening effect for a thin film. Furthermore, the  wavelength of the driving field $\bm{h}$, corresponding to a frequency close to the ferromagnetic resonance  $c/\omega\sim\SI{e-1}{\m}$, is much larger than the characteristic dimension of the typical micrometer-scale beam and, since we do no make particular assumptions on $\bm{h}$, we can write the quasi-static approximation:
\begin{equation}
 \nabla\times\bm{H}_0\sim 0\sim \nabla\times\bm{h},
\end{equation}
provided that we consider the screened $\bm{h}$ in the conducting case.

Next, we derive the quantum mechanical Hamiltonian for the magnon mode assuming a static deformation of the beam.
We express the magnetization as $\bm{M}=\bm{M}_0+\bm{m}$, where $\bm{M}_0$ provides the direction of the static magnetization in the presence of a static external magnetic field $\bm{H}_0$ and $\bm{m}$ is the deviation to be quantized.
By using the assumptions of a single domain magnet and the external field in $yz$ plane,
we have the situation depicted in Fig.~\ref{fig:beamsketch}
\begin{equation}
    \bm{H}_0= H_0 (\cos\phi\, \hat{z}+ \sin\phi\, \hat{y}),\,
    \bm{M}_0= M_S (\cos\theta\, \hat{z}+ \sin\theta\, \hat{y}).
\end{equation}
That is, the magnetization $\bm{M}_0$ is also in the $yz$ plane, meaning that the static deformation of the beam does not remove the hard axis in $x$ direction.
The magnetization angle $\theta$ depends on the direction and size of the field $\bm{H}_0$. 
This dependence results from
the competition between the external, the
magnetoelastic, and the demagnetizing fields, as the angle $\theta$ is determined by the free energy minimum from Eq.~\eqref{eq:averagemagneticF}.
Aside from the global free energy minimum, there can exist a local minimum corresponding to a metastable magnetization configuration.
This describes magnetic hysteresis.
Furthermore, the ferromagnetic
resonance frequency depends on the effective field and, then, on
$\bar{\epsilon}_{zz}$, giving rise to an optomechanics-like coupling. The
magnetic anisotropy related with the strain also provides a coercive field which in the case of
a magnetic field perpendicular to the beam is
$H_c=2|\Bc/(\mu_0 M_S^2)+1|\bar{\epsilon}_{zz}^{(0)} M_S$.
For further details, see Appendix~\ref{sec:magnetic_hysteresis}.

We assume a Kittel-type magnon mode $\bm{m}$ that is spatially uniform in the ferromagnetic beam and neglect the dynamics of finite momentum magnons.
This follows from the magnon driving setup: we assume that the magnetic field $\bm{h}$ is homogeneous in space and, thus, the spatially uniform mode is dominantly excited.
In practice, however, the finite momentum magnons provide an effective thermal bath for the vibrations and the spatially uniform magnon mode.
The spectrum of such a bath would then depend on the external magnetic field.

 The quantum magnetization Hamiltonian can be obtained  by
 substituting the Holstein-Primakoff relations in the total magnetic
 energy $H_{mg} = \int dx^3 \mathcal{F} = L A \bar{\mathcal{F}}$ (see
 Appendix~\ref{sec:magnetic_hamiltonian}). Choosing the reference
   direction $\hat z'$ along the static magnetization $\bm{M}_0$
   ($\hat z'=\cos \theta \, \hat z + \sin \theta \, \hat y$), we can expand and get
\begin{subequations}
\begin{eqnarray}
    && M_{z'}= M_S-\frac{\hbar \gammah}{L A} \hat{m}^\dagger \hat{m},\\
    && M_{+'}= m_{\rm ZPQ} \hat{m},\quad
    M_{-'}=m_{\rm ZPQ} \hat{m}^\dagger,\nonumber\\
\end{eqnarray}
\end{subequations}
where $m_{\rm ZPQ}=\sqrt{\frac{2 \hbar \gammah M_S}{L A}}$ and $\gamma$ is the gyromagnetic ratio.
The quantized magnetization components $M_{z'}$ and $M_{y'} = \frac{1}{2i}(M_{+'} - M_{-'})$ are related to those in the free energy density $\bar{\mathcal{F}}$ in Eq.~\eqref{eq:averagemagneticF} by a rotation of the angle $\theta$ about the $x$ axis which leaves the $x$ component invariant, $M_x = M_{x'} = \frac{1}{2}(M_{+'} + M_{-'})$.

Without considering the driving terms $\bm{h}$ and the coupling with
the flexural dynamics $\epsilon^{(1)}_{zz}$, the magnetic Hamiltonian
reads to the leading order in $1/M_S$
\begin{equation}
\label{eq:baremagneticH}
    \frac{\hat{H}_m}{\hbar} = \omega_{1} \hat{m}^\dagger \hat{m}+
     \frac{\omega_{2}}{2} [\hat{m}^2 +
      (\hat{m}^\dagger)^2] +\omega_3\, i \frac{\hat{m}^\dagger-\hat{m}}{\sqrt{2}},
\end{equation}
where
\begin{subequations}
\begin{eqnarray}
 && \omega_{1}= \gammah \mu_0 \left[ H_0 \cos(\phi-\theta)+ \frac{M_S}{2} \right] \nonumber\\
 &&\qquad - 3\bar{\epsilon}_{zz}^{(0)} \gammah\mu_0 M_S \cos^2\theta,\nonumber\\ 
 &&\qquad - \bar{\epsilon}_{zz}^{(0)}\frac{\gammah\Bc}{M_S} \left( 3 \cos^2\theta-1\right)\\
  && \omega_{2}= \gammah\mu_0\frac{M_S}{2}
 -\bar{\epsilon}_{zz}^{(0)}\frac{\gammah\Bc}{M_S} \sin^2 \theta\nonumber\\
 &&\qquad -  \bar{\epsilon}_{zz}^{(0)} \gammah\mu_0 M_S \left(\sin^2 \theta+1\right),\\
 && \omega_3 = -\frac{\sqrt{2}\gamma M_S}{m_{\rm ZPQ}} \biggl[ \bar{\epsilon}_{zz}^{(0)} \left( \frac{\Bc}{M_S}
 + \mu_0M_S \right) \sin2\theta\nonumber\\
 &&\qquad + \mu_0 H_0 \sin(\phi-\theta)
 \biggr].
\end{eqnarray}
\label{eq:omegas}
\end{subequations}
The term proportional to $\hat{m}^\dagger\, \hat{m}$ is the sum of
four components: one coming from the external magnetic field $H_0$,
the second and third from the hard axis and the last from the magnetoelastic coupling. On the other hand, $\omega_2$ does not depend on the external magnetic field. This is a consequence of the assumption that $\bm{H}_0$ is parallel to the $yz$ plane. 
Finally, the linear term $\omega_3$ does not depend on the $x$-hard axis strength which is quadratic in $M_x$ in  Eq.~\eqref{eq:averagemagneticF}.

$\hat{H}_m$ is a highly tunable quadratic Hamiltonian which could be relevant in the continuous-variable quantum information processing~\cite{serafini2017quantum}: the hard and easy axes produce a controlled squeezing. This leads to entanglement between the spins in the magnetic material \cite{zou2020tuning}.
Recently, it has been suggested that by tuning the frequency $\omega_1$ close to $\omega_2$ and coupling the magnon system to a microwave cavity, a magnon Schrödinger cat state (superposition of two coherent states) could be observed~\cite{sharma2021spin}.
The fact that the magnon Hamiltonian is often not diagonal (i.e. $\hat{H}_m = \hbar\omega_K\hat{m}^\dagger \hat{m}$) should also have other signatures, e.g., in the shot noise of ferromagnet-conductor systems~\cite{kamra2016super}.

Let us diagonalize the Hamiltonian~\eqref{eq:baremagneticH}.
For this task, it is useful to introduce dimensionless quadrature operators $\hat{x}_m = (\hat{m}^\dagger + \hat{m})/\sqrt{2}$ and $\hat{p}_m = i(\hat{m}^\dagger - \hat{m})/\sqrt{2}$ (the subscript indicates the corresponding bosonic mode).
The Hamiltonian~\eqref{eq:baremagneticH} may be expressed as
\begin{align}
    \frac{\hat{H}_m}{\hbar} = \frac{\omega_1 + \omega_2}{2}\hat{x}_m^2 + \frac{\omega_1 - \omega_2}{2} \hat{p}_m^2 + \omega_3 \hat{p}_m.
\end{align}
The diagonalization is achieved by first displacing $\hat{p}_m$ and then scaling both $\hat{x}_m$ and $\hat{p}_m$ properly.
That is, we define a new bosonic operator $\hat{l}$ so that it has the quadrature operators $\hat{p}_l = \sqrt{c} (\hat{p}_m + d)$ and $\hat{x}_l = \hat{x}_m/\sqrt{c}$ where $c = \sqrt{(\omega_1-\omega_2)/(\omega_1 + \omega_2)}$ and $d = \omega_3/(\omega_1 - \omega_2)$.
In the validity range of the Euler-Bernoulli theory $\bar{\epsilon}_{zz}^{(0)} \ll 1$, we generally find that $c < 1$.
Thus, we call $c$ the squeezing factor.
The variance of $\hat{p}_l$ is smaller than the variance of $\hat{p}_m$ by a factor of $c$; the corresponding factor is $1/c > 1$ for the variances of $\hat{x}_l$ and $\hat{x}_m$. 
The diagonalized Hamiltonian is then given by $\hat{H}_m = \hbar\omega_K \hat{l}^\dagger \hat{l}$ 
where the FMR frequency $\omega_K = \sqrt{\omega_1^2 - \omega_2^2}$ may be expressed as
\begin{align} \label{eq:Kittel_freq}
    \omega_K &= \gammah\mu_0
    \left[H_0 \cos(\theta -\phi ) +  \mathcal{M}_{\epsilon} \cos(2 \theta )\right]^{\frac{1}{2}}\\
    &\quad\!\times \left[H_0 \cos(\theta -\phi ) + (1 - 2 \bar{\epsilon}_{zz}^{(0)})M_S + \mathcal{M}_{\epsilon}  \cos^2(\theta)\right]^{\!\frac{1}{2}}. \nonumber
\end{align}
Here, $\mathcal{M}_{\epsilon} = -2 \bar{\epsilon}_{zz}^{(0)}
\left(\frac{\Bc}{\mu_0 M_S^2} + 1\right) M_S=\pm H_c$ describes the
anisotropy energy due to magnetoelastic and demagnetizing field
terms. The sign of this term depends on the relative magnitude of the
two effects in the typical situation with $\Bc < 0$.
The corresponding bosonic transformation is given by
\begin{subequations}
\label{eq:diagonal_transfor}
\begin{eqnarray}
    \hat{l} \, &&= \zeta_+\, \hat{m} + \zeta_-\, \hat{m}^\dagger + i\, \frac{(\zeta_+ + \zeta_-)\, \omega_3}{\sqrt{2}\omega_K},\\
    \zeta_\pm &&=\sqrt{\frac{1}{2}\left( \frac{\omega_1}{\omega_K} \pm 1 \right)}.
\end{eqnarray}
\end{subequations}
Note that without strain, i.e., $\bar{\epsilon}_{zz}^{(0)} = 0$, we obtain $\omega_K = \gammah \mu_0 \sqrt{H_0(H_0 + M_S)}$ as expected while $c = \sqrt{H_0/(H_0 + M_S)}$ and $d = 0$.
Therefore, the squeezing is inherent to the magnon system and produced by the  out-of-plane hard axis while the displacement is due to the static deformation.

\subsection{Magnomechanical coupling}
\label{subs:magnomechanicalcoupling}

By introducing the flexural component of the strain $\bar{\epsilon}_{zz}^{(1)}$ via Eqs.~\eqref{eq:u1} and \eqref{eq:strainexpansion} and replacing $\bar{\epsilon}_{zz}^{(0)}$ in Eq.~\eqref{eq:omegas} by the full $\bar{\epsilon}_{zz}$, we have both a magnomechanical and linear coupling.

\begin{figure}
\centering
{\includegraphics{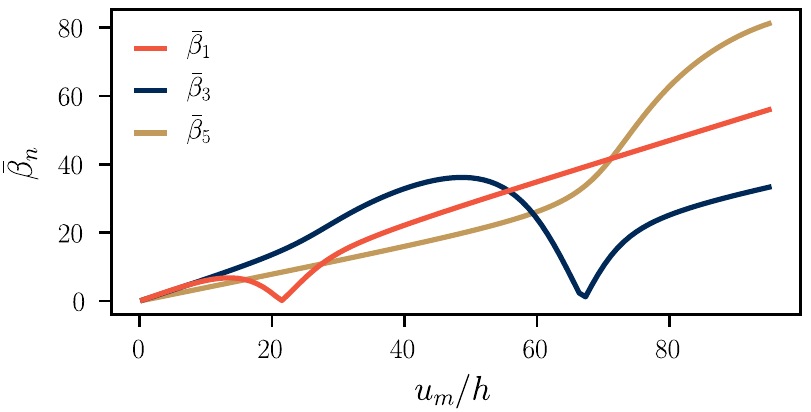}} 
\caption{The mechanical mode parameter $\bar{\beta}_n$ as a function of the static bending parameter $u_m$ divided by the height~$h$ of the beam for the three first $n = \mathrm{odd}$ modes.}
\label{fig:beta1}
\end{figure}

The magnomechanical interaction Hamiltonian reads
\begin{subequations}
\label{eq:magnomechanicalH}
\begin{eqnarray}
     \hat{H}_{mm}&&=\hbar \sum_{n} (g_n^{\text{me}}-g_n^{\text{dm}}) (\hat{b}_n^\dagger+\hat{b}_n) \bigl[ \hat{m}^\dagger \hat{m} (3 \cos^2\theta-1) \nonumber\\
     &&\!\!\!\!+\frac{(\hat{m}^\dagger)^2 + \hat{m}^2}{2} \sin^2\theta 
     -\frac{M_S L A}{\hbar \gamma} \cos^2\theta-\frac{\sin^2\theta}{2}\bigr]\\
    &&\!\!\!\!-\hbar \sum_{n} g_n^{\text{dm}} (\hat{b}_n^\dagger+\hat{b}_n) \bigl[ \hat{m}^\dagger \hat{m} +\frac{(\hat{m}^\dagger)^2 + \hat{m}^2}{2} +1\bigr],\nonumber\\
     g_n^{\text{me}}&&=-\bar{\beta}_n \frac{h x^{\rm ZPM}_n}{ L^2}  \frac{\gammah\Bc}{M_S},     \\
     g_n^{\text{dm}}&&=\bar{\beta}_n \frac{h x^{\rm
                       ZPM}_n}{L^2} 
                       \gammah \mu_0 M_S,      \\
    \bar{\beta}_n &&= \frac{L}{h} \int_0^L d z (\partial_z u_0) (\partial_z \chi_n),
\end{eqnarray}
\end{subequations}
where $g_n^{\text{me}}$ and $g_n^{\text{dm}}$ refer to magnetoelastic
and demagnetizing coupling respectively. From the dimensionless mode parameters
$\bar{\beta}_n$ we get a selection rule: only the modes with
odd $n=1,3,\dots$ corresponding to even mode functions  give rise to a
coupling.  
In Fig.~\ref{fig:beta1}, we plot  $\bar{\beta}_n$ for the three lowest even modes as a
function of the static bending parameter $u_m$.
We note that the change in the eigenmode shape $\chi_n$ is as a function of $u_m$ so drastic that, at certain points, $\chi_n$ becomes orthogonal to the static deformation $u_0$ (for further details, see Appendix~\ref{sec:mechanical_eigenmodes}).
At these points, the magnomechanical coupling vanishes.
Also, rather surprisingly, there is no clear hierarchy of the parameters $\bar{\beta}_n$ but rather, they depend strongly on the static deformation.
It may hence be that a higher-frequency mode couples stronger to the magnetization dynamics.

There is also a linear coupling,
\begin{subequations}
\label{eq:linearcouplingH}
\begin{eqnarray}
    && \hat{H}_l=\hbar \sum_n g_n^{(l)}\, i \frac{\hat{m}^\dagger-\hat{m}}{\sqrt{2}}\,
    \frac{\hat{b}_n^\dagger+\hat{b}_n}{\sqrt{2}} \sin 2 \theta,\\
    && g_n^{(l)}=\bar{\beta}_n \frac{h x_{\rm ZPQ}}{\sqrt{12} L^2}\, \frac{M_S}{m_{\rm ZPQ}}  \gamma \mu_0 {\cal M}_\epsilon.
\end{eqnarray}
\end{subequations}
It also depends on the mechanical mode parameter $\bar{\beta}_n$ and 
is directly proportional to the coercive field $H_c=|{\cal M}_\epsilon|$ measured perpendicular to the beam.
In this work, however, we concentrate on a situation in which the FMR and mechanical frequencies are strongly detuned and the linear coupling will therefore play no further role.

Note that the interaction terms in Eqs.~\eqref{eq:magnomechanicalH} and~\eqref{eq:linearcouplingH} are written in the original non-diagonalized basis, and
  therefore are not yet the relevant ones for magnomechanical
  measurements. Using the transformation \eqref{eq:diagonal_transfor}
  allows us to write this in the diagonal basis. Neglecting doubly
  rotating terms, the result for the magnomechanical coupling Hamiltonian is $\hat{H}_{mm}= \hbar\sum_n g_{m,n} (\hat b_n^\dagger+\hat b_n) \hat l^\dagger \hat l$ with
  \begin{equation}
    g_{m,n}=2 (g_n^{\rm me}-g_n^{\rm dm})\left(c\cos^2 \theta +\frac{1}{c}\cos 2\theta \right) -2c g_n^{\rm dm}. 
    \label{eq:magnomech}
  \end{equation}
  The squeezing factor 
  $$c=\sqrt{\frac{H_0 \cos(\theta -\phi ) +  \mathcal{M}_{\epsilon} \cos(2 \theta )}{H_0 \cos(\theta -\phi ) + (1 - 2 \bar{\epsilon}_{zz}^{(0)})M_S + \mathcal{M}_{\epsilon}  \cos^2(\theta)}}
  $$ depends on the direction of the magnetic
  field in a non-trivial way which is why the magnomechanical
  coupling depends not only on the direction of the field but also its
  precise magnitude setting the direction of the
  magnetization. Moreover, the
  magnitude is different for the stable and metastable magnetizations 
  for fields below the coercive field.

\begin{figure}
    \centering
    \includegraphics{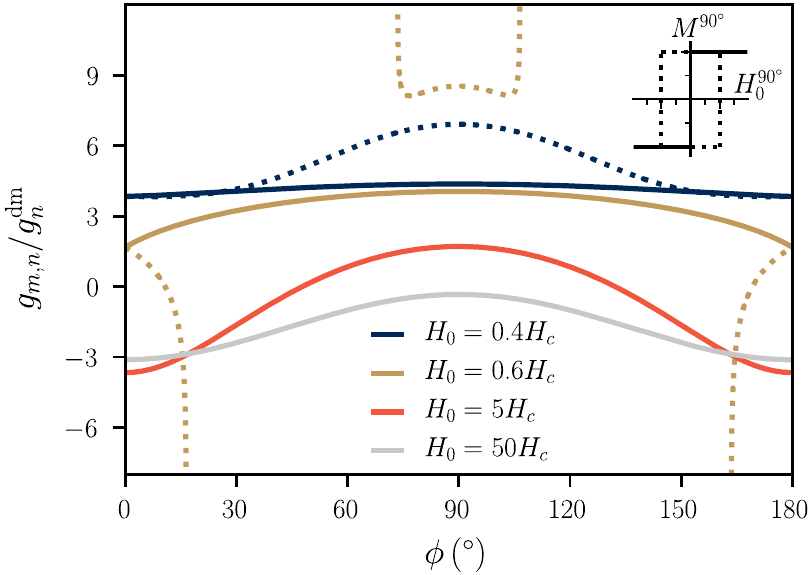}
    \caption{Magnomechanical coupling constant $g_{m,n}$ as a function of the external field direction $\phi$. The different curves represent different magnetic field strengths $H_0$, characterized by the coercive field $H_c$ at $\phi = 90^\circ$: the dotted lines correspond to the metastable magnetization configuration when $H_0 < H_c$. Note that this metastable state does not necessarily exist for all $\phi$. Here, we set $u_m/L = 0.1$ and $B_1 = -0.6 \mu_0 M_S^2$ so that $50 H_c \approx M_S$. The inset shows the corresponding hysteresis curve for $\phi = 90^\circ$.}
    \label{fig:magnomech:zhard}
\end{figure}

We illustrate the dependence of the magnomechanical coupling on the magnetic field direction in Fig.~\ref{fig:magnomech:zhard}, assuming that $z$-axis is a hard axis ($|\Bc| < \mu_0 M_S^2$).
We observe that, in general, as the magnitude of the external field $H_0$ is increased, the magnomechanical coupling decreases.
Focusing on the angle $\phi = 90^\circ$, the magnomechanical coupling changes sign as a function of the external magnetic field strength $H_0$ and approaches the value $g_{m,n} \rightarrow -2 g_n^\mathrm{me}$ as $M_S/H_0 \rightarrow 0$.
Lastly, we find that $H_0 = H_c/2$ represents the point where the metastable magnetization ceases to exist for some values of $\phi$.
For external fields $H_0$ that are between $H_c/2$ and $H_c$ this shows up in the magnomechanical coupling as divergences; it should be noted, however, that the free energy barrier is very small at these points.

\begin{figure}
    \centering
    \includegraphics{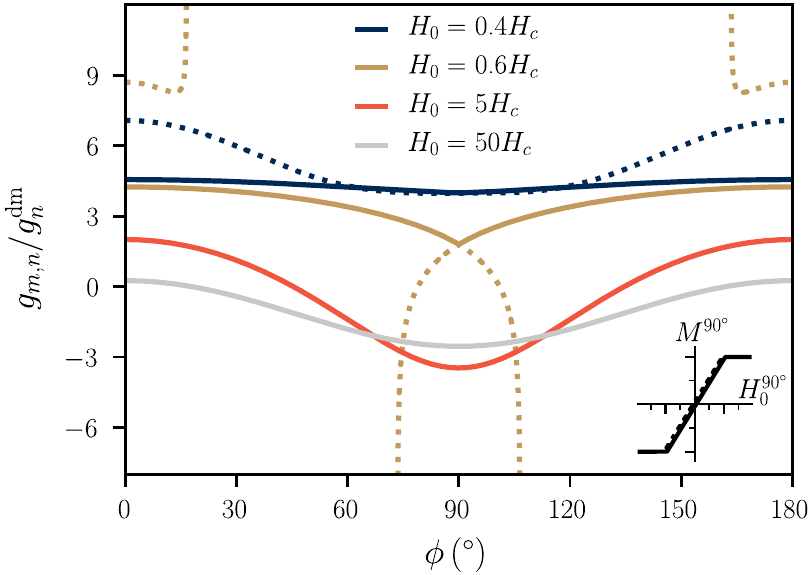}
    \caption{Magnomechanical coupling constant $g_{m,n}$ as a function of the external field direction $\phi$ that is otherwise identical to Fig.~\ref{fig:magnomech:zhard} except we set $B_1 = -1.4\mu_0 M_S^2$. With this value, the coercive field at $90^\circ$ remains the same but $z$-axis becomes an easy axis, changing both the hysteresis curve as well as the magnomechanical coupling. }
    \label{fig:magnomech:zeasy}
\end{figure}

An interesting consequence of our choices for the magnetization free energy is that there are always two values of $B_1$ that correspond to a given coercive field $H_c$.
This changes the magnomechanical coupling constant as shown in Fig.~\ref{fig:magnomech:zeasy}.
There, the magnetoelastic energy is negative ($B_1 < 0$) and dominates, that is, $|B_1| > \mu_0 M_S^2$.
At the same time, the $z$-axis becomes an easy axis.
Thus, a hysteresis measurement separates these two situations with the same coercive field; see the insets of both Figs.~\ref{fig:magnomech:zhard} and~\ref{fig:magnomech:zeasy} which describe the hysteresis curves obtained for an external magnetic field pointing perpendicular to the beam at $\phi = 90^\circ$.

Finally, we note that interesting physics remains even if the magnomechanical coupling vanishes, $g_{m,n} = 0$.
In such a case, a second order cross-Kerr-type term, i.e., a term quadratic in the deformation, may remain finite.
The situation is analogous to the "membrane in the middle" optomechanics setup with a vibrating semitransparent mirror in the center of the cavity which is proposed as a way of measuring the phonon number in the mechanical mode~\cite{thompson2008strong}.
In the doubly clamped magnetic beam, such terms are always present as discussed below Eq.~\eqref{eq:averagemagneticF}.
Interestingly, they can be finite without ($u_0 = 0$) and with the static deformation.
The strength of the cross-Kerr terms may also depend on the external field $\bm{H}_0$.
Especially in the beam with a static deformation, there are multiple values for $H_0$ and $\phi$ which give $g_{m,n} = 0$ as shown in Figs.~\ref{fig:magnomech:zhard} and~\ref{fig:magnomech:zeasy}
However, the analysis of the cross-Kerr terms is beyond the scope of this work.

\section{Driving individual resonances}
\label{sec:driving}

In order to experimentally see the magnomechanical coupling and its effects, it is necessary to consider the ways the magnomechanical system may be driven.
At the same time, dissipation to the environment is still to be taken into account.
Here, we discuss some general features of the driving while showing an example of a possible scheme together with a model for its description using the input-output formalism~\cite{gardiner1985input}.
For brevity, we focus only on reflection measurements.

The driving of the magnon mode $\hat m$ is achieved by applying an alternating magnetic field $\bm{h}$ that is perpendicular to the static field $\bm{H}_0$.
Here, it is convenient to choose $\bm{h}$ in the direction of the displacement $u$, i.e., in the $x$-direction.
Then, $\bm{h}$ is independent of the direction $\phi$ of the static field $\bm{H}_0$.
This produces an extra term in the Hamiltonian, proportional to $\mu_0\lvert\bm{h}\rvert M_x$ or $\lvert\bm{h}\rvert(\hat m + {\hat m}^\dagger)$.
The coupling rate of a similar drive is described and measured in Ref.~\cite{mckenziesell2019low}; importantly, it is directly proportional to the spatial overlap of the magnon mode and the driving field.

Within the input-output formalism, we do not include such driving to the closed system Hamiltonian, but rather assume that there exists a bath of free electromagnetic modes to which the magnon mode $\hat m$ is coupled.
Integrating out these free modes gives rise to dissipation and the possibility of the bath exciting the system, either via thermal noise or an external drive.
In the diagonalized frame, the squeezing factor $c$ is introduced to the coupling between the bath and the magnon which modifies the dissipation rate.

Since the alternating magnetic field can be produced by an alternating current, we can readily associate the input and output fields of the magnon mode to in- and outgoing voltage signals on a transmission line.
If we now denote the driving field by $\hat{l}_\mathrm{in}$, we have the input-output relation $\hat{l}_\mathrm{out} - \hat{l}_\mathrm{in} = \sqrt{\kappa_e} \hat l$ in the diagonalized frame.
The dynamics is then determined by
\begin{align}
    \dot {\hat l} = \frac{i}{\hbar} [\hat{H}_S, \hat l] - \frac{\kappa}{2} \hat l - \sqrt{\kappa_e} \hat{l}_\mathrm{in},
    \label{eq:io:magnon}
\end{align}
where $\hat{H}_S = \hat{H}_m + \hat{H}_{mm} + \hat{H}_{mc}$ is the system Hamiltonian, $\kappa$ the total effective magnon linewidth, and $\kappa_e$ the dissipation rate related to external driving.
In the proposed setup, we expect that $\kappa_e/\kappa \ll 1$ due to only a small proportion of the external field residing within the beam.

Similar to traditional optomechanics, the mechanical mode does not necessarily need to be actuated. However, it may still be driven and characterized, for instance, within the magnetomotive scheme \cite{cleland1996fabrication,Yurke1996Magnetomot,Pashkin08} by utilizing the static field $\bm{H}_0$.
If an alternating current $I_d$ goes through the beam, a Lorentz force of magnitude $f =  I_d \mu_0(H_0 \sin{\phi} + M_S \sin{\theta})$ per unit length acts on the beam to the direction $\hat n(z)$ that is locally perpendicular to the beam as in Fig.~\ref{fig:beamsketch}b.
We may neglect the dynamical magnetization in this force as we assume its characteristic frequency to be much larger than those of the vibrational eigenmodes.
Likewise, the flexural (dynamic vibrational) modes should have a negligible effect on the direction $\hat n(z)$ and, thus, we can consider only the normal of the static deformation $u_0$ here.
It is possible to find the $x$-component of the force, $f_x(z)$, which in turn must be projected to the eigenmodes $\chi_n$ to determine its effect on the amplitude $x_n$.
The effective force per unit length on $x_n$ is
\begin{align}
    f_\mathrm{eff} =  \int_0^L \frac{dz}{L} \chi_n f_x = f\int_0^L \frac{dz}{L} \frac{\chi_n}{\sqrt{1 + (\partial_z u_0)^2}}.
\end{align}
It is possible redefine the mass $m$ in Eq.~\eqref{eq:mechanicalH} to contain the integral term; this gives the so-called effective mass~\cite{aspelmeyer2014cavity}.
Due to the mirror symmetry of the static beam, only the symmetric vibrational eigenmodes couple to this force.
Note that a similar analysis holds for the force in the $z$ direction: such force $f_z$ is negligible when $\partial_z u_0 \ll 1$ and it only couples to the antisymmetric vibrational eigenmodes.
By changing the frequency of the drive $I_d$, we may then drive individual flexural modes on resonance if the vibrations are of high quality.

What is then observed is the induced voltage, as the beam vibrates and thus changes a magnetic flux through a circuit in the $xz$-plane.
This induced voltage is
\begin{equation}
    V_{emf} = \mu_0 H_0 \sin(\phi) \int_0^L d z \frac{\partial u(z)}{\partial t} = \mu_0 H_0 \sin(\phi) L \sum_n \dot{x}_n.
    \label{eq:inducedV}
\end{equation}
That is, all flexural modes contribute to the voltage.
The static magnetization $\bm{M_0}$ may be neglected as it does not provide a change in the magnetic flux in the first order while the dynamical magnetization changes at much higher frequency than the vibrations.

The modeling of the magnetomotive scheme can be done within the input-output formalism.
First, we define formally the input and output fields for each vibration mode $\hat{b}_n$ by the usual relation $\hat{b}_{n,\mathrm{out}} - \hat{b}_{n,\mathrm{in}} = \sqrt{\gamma_{n,e}} \hat{b}_n$ where $\gamma_n$  refers to the total dissipation rate of the vibration mode $n$ whereas $\gamma_{n,e}$ refers to external losses.
One may then transform this formal input-output relation to match the induced voltage in Eq.~\eqref{eq:inducedV} by multiplying by $H_0 L\sin{\phi}$, using $\hat{x}_n \propto (\hat{b}_n + \hat{b}_n^\dagger)$, and taking the time derivative.
These transformations also define new input and output fields.
However, in the Fourier space, these are linear transformations so we may equally well consider the original input-output relation, as long as we remember that all the fields are in fact proportional to $\sin{\phi}$ and that $\gamma_{n,e}$ depends on the force projection on the eigenmodes.
The dynamics of the flexural modes are obtained by
\begin{equation}
    \dot {\hat b}_n = \frac{i}{\hbar} [\hat{H}_S, \hat{b}_n] - \frac{\gamma_n}{2} \hat{b}_n - \sqrt{\gamma_{n,e}} \hat{b}_{n,\mathrm{in}}
    \label{eq:io:vibrations}
\end{equation}
from which it is straightforward to obtain the output $\hat{b}_{n,\mathrm{out}}$.

Lastly, it should be noted that there is ''cross-driving'', i.e., the different drives exemplified here interact if they are applied simultaneously.
The alternating current $I_d$ through the beam causes a force proportional to $\mu_0 \lvert\bm{h}\rvert I_d$ to the beam.
Likewise, it causes an extraneous magnetic flux which generates induced voltage proportional to the magnitude of current $I_d$.
These effects are negligible in most cases with a frequency mismatch of the magnon and flexural modes.

\section{Magnomechanics}

With the magnomechanical Hamiltonian and the scheme to drive and observe such a system, we may describe an example magnomechanics experiment.
The wealth of literature on optomechanical systems can be used straightforwardly due to the similar form of the Hamiltonian.
However, there are a few issues that are important for magnomechanics specifically.

The relevant magnitude of the eigenfrequencies and dissipation rates of the magnon and mechanical system affect the possible measurements.
As mentioned in the derivation of the magnomechanical Hamiltonian, we assume that the mechanical frequency is much smaller than the FMR frequency, $\omega_m \ll \omega_K$.
These systems are also assumed underdamped, meaning that for each system the eigenfrequency is larger than the dissipation rate, $\omega_m > \gamma$ and $\omega_K > \kappa$.
The only question left is thus whether the mechanical eigenfrequency $\omega_m$ is larger than the FMR linewidth $\kappa$ or not.
The case $\omega_m > \kappa$ is called the resolved sideband regime which, in optomechanics, has allowed for ground state cooling of the mechanical oscillator~\cite{Teufel2011b,AspelmeyerCool11,marquardt2007quantum} and amplification of microwave signals at the quantum limit~\cite{MechAmpPaper}.
This is also the regime of experiments in Refs.~\onlinecite{zhang2016cavity,potts2021dynamical} performed with microwave cavities and YIG spheres.
However, for many other ferromagnetic materials, it may be expected that $\omega_m < \kappa$. 
Although the analysis could proceed either way, we focus on results in this non-resolved sideband regime where the FMR linewidth $\kappa$ dominates.

Next, we describe in detail an amplification scheme for microwave signals in the non-resolved sideband regime which is both theoretically known~\cite{botter2012linear} and experimentally observed~\cite{cohen2020optomechanical} in optomechanical systems.
Especially, we focus on the aspect that is not present in optomechanics: the tunability of the magnomechanical coupling as well as the magnon eigenfrequency with respect to the external static field $\bm{H}_0$.

The derivation of the reflection coefficient, often called $S_{11}$, follows the lines of Ref.~\onlinecite{botter2012linear}.
First, we assume a strong drive on the magnon system at a frequency $\omega_d$ and focus on the deviations around the driven system. That is, we replace $\hat{l} \rightarrow (\sqrt{n} + \hat{l})e^{- i \omega_d t}$, where we identify $n$ as the number of magnons, in the Hamiltonians of Eqs.~\eqref{eq:baremagneticH},~\eqref{eq:magnomechanicalH}, and~\eqref{eq:linearcouplingH}, using the transformation in Eq.~\eqref{eq:diagonal_transfor}.
Note that the Holstein-Primakoff transformation gives an upper limit to the strength of the drive, given by $M_S \gg \hbar \gammah n/(LA)$.
Then, we may neglect the second order terms of the deviations as well as the terms that rotate at frequency $\omega_d$ or faster.
The dynamical part of the Hamiltonian then reads in the diagonalized frame
\begin{equation}
    \hat{H}_S/\hbar = \omega_K \hat{l}^\dagger \hat{l} + \sum_j \omega_j \hat{b}_j^\dagger \hat{b}_j + \sum_j G_j \hat{x}_l \hat{x}_{b,j},
    \label{eq:linearizedHmm}
\end{equation}
where $\hat{x}_{b,j} = (\hat{b}_j + \hat{b}_j^\dagger)/\sqrt{2}$ and the effective coupling constant follows from Eq.~\eqref{eq:magnomech} by $G_j = 2 \sqrt{n} g_{m,j}$, that is,
\begin{equation}
    G_j = 4\sqrt{n}
    \left[ \left(g_j^\text{me}-g_j^\text{dm}\right)
    \left(c\cos^2{\theta} + \frac{1}{c} \cos{2\theta} \right) 
    -  c g_j^\text{dm}\right].
    \label{eq:eff_mm_coupling}
\end{equation}
The effective magnomechanical coupling is thus enhanced by the number $n$ of magnons and is tunable by the external field $\bm{H}_0$, as it determines the direction of the magnetization $\theta$ and the squeezing factor~$c$.

The response matrix of the magnomechanical system is obtained by utilizing the input-output equations in the Fourier space and the transformation between the bosonic operators $\hat{l}$ and $\hat{b}_j$ and their respective quadratures.
For simplicity, let us focus on a single flexural mode $j$ and denote $\omega_j = \omega_m$ while dropping the other subscripts.
We find now the linear response
\begin{equation}
    \begin{pmatrix} \hat{l}(\omega) \\ \hat{l}^\dagger(\omega) \\ \hat{b}(\omega) \\ \hat{b}^\dagger(\omega)\end{pmatrix}
    = T \begin{pmatrix} \hat{l}_\mathrm{in}(\omega) \\ \hat{l}_\mathrm{in}^\dagger(\omega) \\ \hat{b}_\mathrm{in}(\omega) \\ \hat{b}_\mathrm{in}^\dagger(\omega)\end{pmatrix},
\end{equation}
where all elements of the matrix $T$ may be non-zero (further details given in Appendix~\ref{sec:inputoutputdetails}).
Here, it should be noted that $\hat{l}$ is now defined in a frame that corotates with the drive so the frequencies are defined with respect to the drive frequency $\omega_d$.
For example, $\hat{l}_\mathrm{in}(\omega_p)$ corresponds to an input at frequency $\omega_d + \omega_p$ whereas $\hat{l}_\mathrm{in}^\dagger(\omega_p)$ corresponds to $\omega_d - \omega_p$.

The output fields are readily obtained from the relations $\hat{l}_\mathrm{out} = \hat{l}_\mathrm{in} + \sqrt{\kappa_e} \hat{l}$ and $\hat{b}_\mathrm{out} = \hat{b}_\mathrm{in} + \sqrt{\gamma_e} \hat{b}$.
Especially, we can define now the reflection coefficients for the magnon and mechanical systems as $S_{11}^l = 1 + \sqrt{\kappa_e}T_{11}$ and $S_{11}^b = 1 + \sqrt{\gamma_e} T_{33}$.
Likewise, the transduction coefficients may be defined as $S_{lb} = \sqrt{\kappa_e}T_{13}$ (from mechanics to magnons) and $S_{bl} = \sqrt{\gamma_e}T_{31}$ (vice versa).

The amplification of the probe signals may be observed by calculating the reflection coefficient.
Assuming that the magnon dissipation rate $\kappa$ is large compared to $\omega_m$ and $\gamma$, we obtain for $\omega_d = \omega_K$ and $\omega = \pm \omega_m$
\begin{equation}
    S_{11}^l =  1 - 2 \frac{\kappa_e}{\kappa} \left(1 \pm \frac{G^2}{\kappa \gamma} \right)
\end{equation}
for high quality vibrations $\omega_m \gg \gamma$.
Thus, if the coupling~$G$ is large enough, we find $\lvert S_{11}^l \rvert \gg 1$.
At the same time, the mechanical response is sharply peaked at approximately 
$\tilde{\omega}_m \approx \omega_m - \mathrm{Re}\, \Lambda$ where
\begin{equation}
    \Lambda = \frac{i \omega_m \Delta G^2}{[i (\tilde{\omega}_m+ \omega_m) - \frac{\gamma}{2}][\Delta^2 + (i \tilde{\omega}_m - \frac{\kappa}{2})^2]}
\end{equation}
and $\Delta = \omega_K - \omega_d$.
This change in the resonant frequency corresponds to the optical spring effect of optomechanics.
At this frequency, the reflection coefficient for the force driving the mechanics is given by $S_{11}^b = 1 - \gamma_e/(\frac{\gamma}{2} + \mathrm{Im}\, \Lambda)$. 
Thus, the current through the beam may be modified by the magnomechanical coupling.

\begin{figure}
    \centering
    \includegraphics{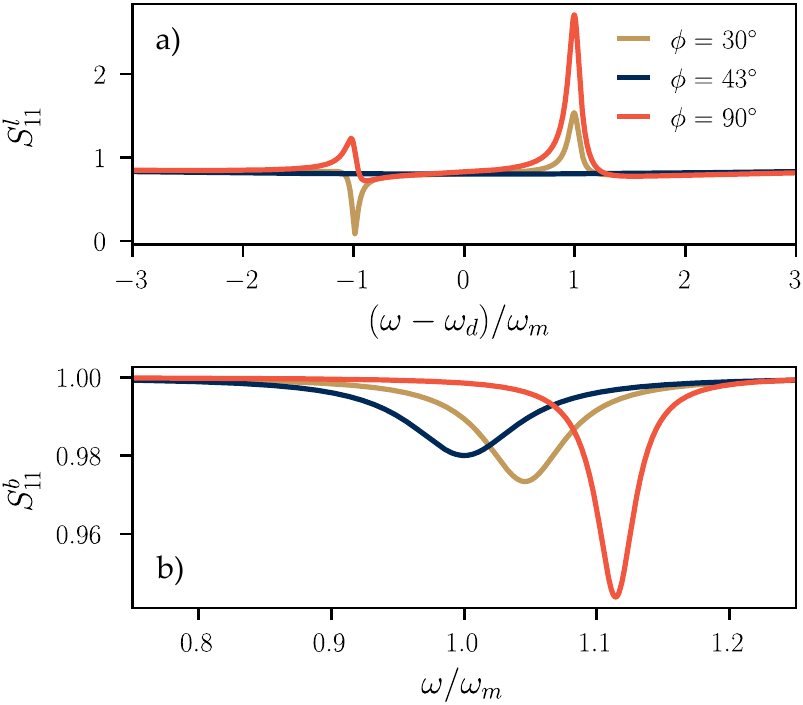}
    \caption{Reflection coefficients of the magnon a) and flexural b) systems for three different angles $\phi$ while the magnitude of the external field $H_0 = 5 H_c \approx 0.1 M_S$ is fixed. 
    In a), we set $\Delta = 0$ whereas in b) $\Delta = - \omega_m$ for all angles. 
    The magnitude of the effective coupling rate is given by $2 \sqrt{n}g_j^\mathrm{dm} = 0.8\omega_m$, and the dissipation rates are $\kappa = 15\omega_m$, $\gamma = 0.1\omega_m$, $\kappa_e/\kappa = 0.1$, and $\gamma_e/\gamma = 0.01.$ 
    We choose here relatively low quality vibrations for visual clarity.
    Otherwise, the parameters match those of Fig.~\ref{fig:magnomech:zhard} so that the orange line there characterizes the angular dependence of the magnomechanical coupling constant.
    }
    \label{fig:rc_vs_freq}
\end{figure}

The reflection coefficients for different angles $\phi$ of the external field $\bm{H}_0$ are graphed in Fig.~\ref{fig:rc_vs_freq}. 
With the chosen parameters, the effective coupling constant~\eqref{eq:eff_mm_coupling} vanishes at $\phi \approx 43^\circ$ (see also the orange line in Fig.~\ref{fig:magnomech:zhard}) and, thus, the responses match those of the uncoupled systems (blue line).
It would be possible to obtain larger values of the coupling constant with values $\phi \approx 0$ but, as discussed in Sec.~\ref{sec:driving}, this would not be commensurate with the magnetomotive scheme for which the input and output amplitudes depend on $\sin\phi$.

\section{Experiment on a magnetic beam}

\begin{figure}
\centering
{\includegraphics[width= \columnwidth]{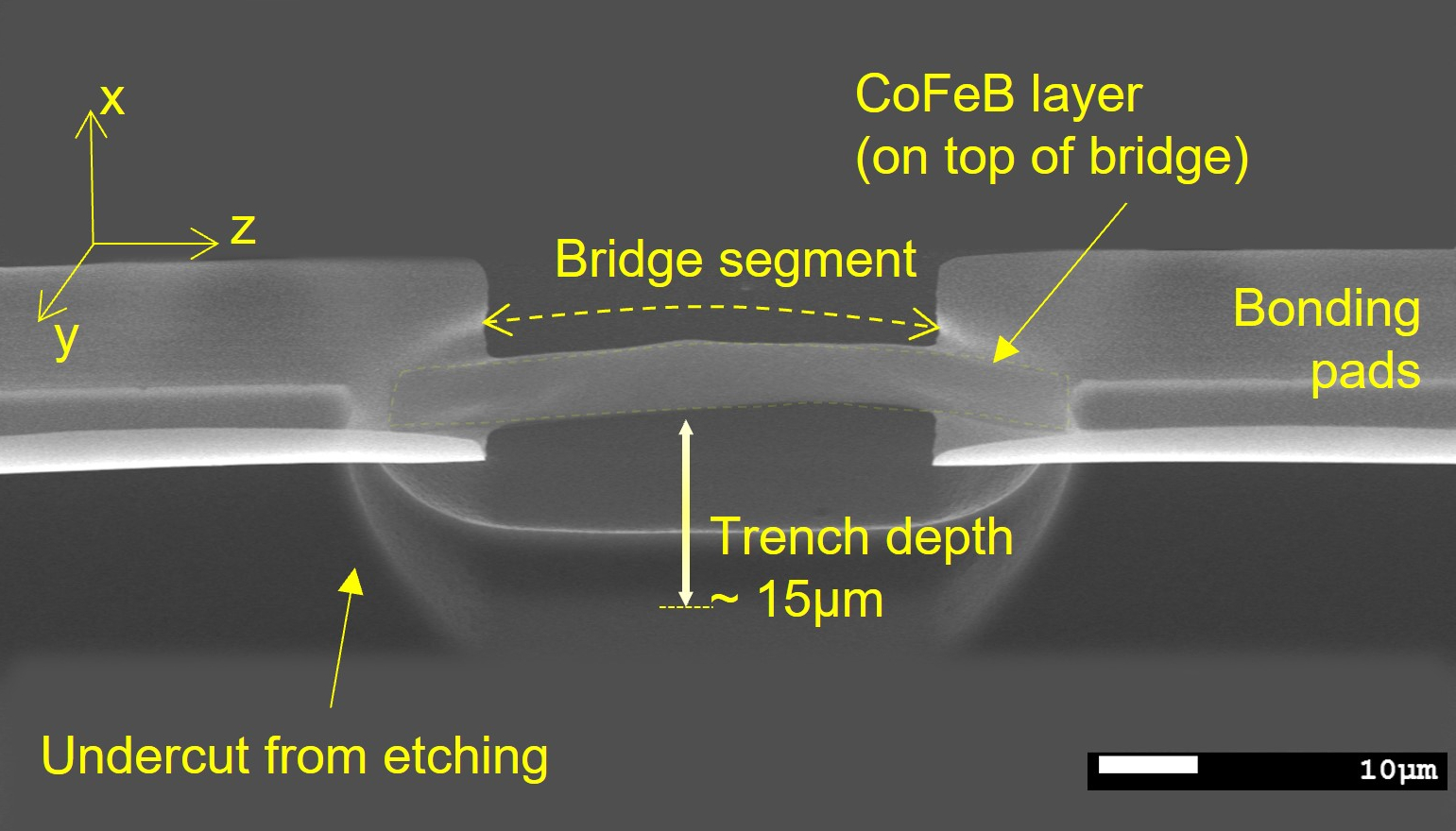}} 
\caption{Scanning electron micrograph of a doubly clamped beam fabricated in our laboratory (see main text for further details). A close inspection of the beam or bridge segment reveals that the beam is deformed but its structure differs slightly from the idealized beam in Fig.~\ref{fig:beamsketch}. The process of reactive ion etching during the fabrication also leads to the formation of an undercut, further contributing to the asymmetry.}
\label{fig:sample}
\end{figure}

Already in some experiments, magnetic beams and cantilevers have been considered \cite{Freeman010MagCantil,Arisawa2019deltaE,Schmidt2019YIGbridge}. Here we demonstrate successful fabrication and characterization of a magnetic beam made out of CoFeB, which exhibits a large magnetostriction, and is promising for the upcoming magnomechanics experiments.
The beam is a bilayer system, consisting of non-magnetic aluminium and the CoFeB layer.
Using a combination of electron-beam lithography, evaporation, DC sputtering and a lift-off-based process, we first fabricated the beam structures over a silicon substrate. Subsequently, using reactive ion etching,
we created the suspended bridge structures. The final fabricated bridges had a length of \SI{50}{\micro\metre}
and a width of \SI{10}{\micro\metre}. The aluminium layer had a thickness of \SI{100}{\nano\metre} while the CoFeB layer was \SI{50}{\nano\metre} thick (with a Ta capping layer of \SI{3}{\nano\metre}). Consequent to the fabrication steps, an asymmetric deformation is observed in the bridge at room temperature as seen in Fig.~\ref{fig:sample}. This deformation may be attributed to a compressive stress owing to a mismatch of the elastic constants in the different layers. Since the thermal expansion coefficients of these layers are also different, it is expected that the compressive stresses observed at room temperature may be changed at cryogenic temperatures. Nevertheless, as pointed out in the above sections, the presence of the asymmetric deformation is necessary for providing a sizable magnomechanical coupling.

We also characterized the magnetic hysteresis of the suspended beam and found the low-temperature switching field $H_{\rm sw}= \SI{30}{mT}$ in the presence of a field in the $y$ direction, i.e., perpendicular to the wire. This hence corresponds to the case considered in Fig.~\ref{fig:magnomech:zhard}. Using the saturation magnetization of CoFeB, $M_S \approx \SI{1200}{emu/cm^3}$  corresponding to $\mu_0 M_S \approx \SI{1.5}{T}$ or $\gamma \mu_0 M_S \approx \SI{42}{GHz}$, this $H_{\rm sw}$ would be obtained with $\bar{\epsilon}_{zz}^{(0)}[1+B_1/(\mu_0 M_S^2)] = 0.01$ assuming that $\lvert B_1\rvert \ll \mu_0 M_S^2$. This corresponds to the maximum deformation $u_m \approx 0.06 L$, quite well in line with what is seen in Fig.~\ref{fig:sample}.
It should, however, be noted that the magnetoelastic constant $B_1$ does not only depend on the material but also on the details of fabrication~\cite{sander1999correlation,gowtham2016thickness}. 
To our knowledge, there are no accurate estimates available in the literature that would correspond to our system.

The mechanical eigenfrequency may be estimated by using the solved dimensionless eigenfrequencies in Fig.~\ref{fig:mech_eigenfr} (see Appendix~\ref{sec:mechanical_eigenmodes}) and, for simplicity, by using the characteristic values for a beam fully made of aluminum, $\rho = \SI{2.7}{g/cm^3}$ and $E = \SI{70}{GPa}$.
Since $u_m/h \approx 20$ according to the analysis based on the switching field, we find that the eigenfrequency of the first mode is $\omega_1/(2\pi) \approx \SI{1}{MHz}$.
This corresponds to the zero-point motion amplitude $x_1^\mathrm{ZPM} \approx \SI{6e-15}{m}$.
In contrast, if $u_m/h = 10$, we would find the eigenfrequency $\SI{0.5}{MHz}$ as the conversion from Fig.~\ref{fig:mech_eigenfr} is given by $\omega_n \approx (\SI{88}{kHz})\times\bar{\omega}_n$.

Finally,  let us estimate the expected size of the  magnomechanical coupling for this setup. 
Using $u_m/h \approx 20$, which gives the mechanical mode parameter $\bar{\beta}_1 \approx 3$, we get $g_1^\mathrm{dm} \approx \SI{50}{mHz}$.
This is the scale used in Figs.~\ref{fig:magnomech:zhard} and \ref{fig:magnomech:zeasy} and depending on the precise magnetic field the actual magnomechanical coupling may be somewhat larger.
Due to the non-linear nature of the Euler-Bernoulli equation used, if the static deformation was half as large, $u_m/h = 10$, we would find $g_1^\mathrm{dm} \approx \SI{0.15}{Hz}$.
These couplings are comparable to what was found for the magnomechanical coupling in \cite{zhang2016cavity,potts2021dynamical} for YIG spheres and somewhat smaller than the optomechanical coupling in our setups (e.g., $g \sim 100$ Hz in \cite{CasparAmp}).

Experimental preparations on realizing the magnomechanical physics are currently ongoing in our laboratory.

\section{Conclusions}

We present a detailed analysis of magnetoelastic interactions in suspended micromechanical beams made out of ferromagnetic materials. We find the mechanical vibrations of the beam, and its ferromagnetic resonance, exhibit nonlinear interactions reminiscent of radiation-pressure coupling in cavity optomechanics or in microwave optomechanics. The interaction, however, is more versatile and easily configurable. Part of the interaction arises via magnetoelasticity, where the vibrations modulate the frequency of the magnetic resonance. The dominant coupling under typical conditions, however, is due to demagnetizing field of the beam, which is affected by the instantaneous shape of the vibrating beam. The predicted radiation-pressure coupling rates are smaller than in microwave optomechanics, but still sizable enough that optomechanical physics such as cooling, amplification and lasing are within experimental reach. In comparison to optomechanics, our system has the assets of a small footprint, a reconfigurable interaction, and a high power tolerance of the magnon resonance.

\begin{acknowledgments} 
We thank Rasmus Holländer, Huajun Qin and Sebastiaan van Dijken for useful discussions.
This work was supported by the Academy of Finland (contracts 307757, 312057, 317118, 321981), by the European Research Council (contract 615755), and by the Centre for Quantum Engineering at Aalto University.
K.S.U.K. acknowledges the financial support of the Magnus Ehrnrooth foundation.
We acknowledge funding from the European Union's Horizon 2020 research and innovation program under grant agreement No.~732894 (FETPRO HOT). We acknowledge the facilities and technical support of Otaniemi research infrastructure for Micro and Nanotechnologies (OtaNano) that is part of the European Microkelvin Platform.
\end{acknowledgments}

\appendix

\section{Beam dynamics}

\label{sec:mechanical_eigenmodes}

From the non-linear Euler-Bernoulli equation~\eqref{eq:non-linear_EB} one can find a beam configuration that has a static deformation as well as the eigenmodes in which the beam vibrates~\cite{nayfeh2008exact}.
These eigenmodes depend directly on the static deformation which is caused in our description by a negative tension $T_0$.
That is, there is a constant compressive stress or ''load'' which we denote by $P = -T_0 > 0.$

For the following analysis, it is useful
to make Eq.~\eqref{eq:non-linear_EB} dimensionless by introducing the relations
\begin{subequations}
\label{eq:dimensionless_converter}
\begin{eqnarray}
 &&\bar{u}= \sqrt{\frac{A}{I_x}} u, \quad \bar{z}=\frac{z}{L},\quad
 \bar{t}= \sqrt{\frac{E I_x}{\rho A L^4}}\, t,\\
 &&
  \bar{T}_0 = \frac{L^2}{E I_x} T_0, \quad \bar{f} = \frac{A^{1/2} L^4}{E I_x^{3/2}} f,
\end{eqnarray}
\end{subequations}
where $f = f(z)$ is the external force per unit length on the beam.
The dimensionless non-linear Euler-Bernoulli equation reads ($\bar{P} = -\bar{T}_0$)
\begin{align}\label{eq:dimensionlessEB}
     \frac{\partial^2 \bar u}{\partial \bar{t}^2} + \frac{\partial^4 \bar{u}}{\partial \bar{z}^4} 
 + \left[\bar{P} - \frac{1}{2} \int_0^1 d\bar z \left( \frac{\partial \bar u}{\partial \bar z} \right)^2  \right] \frac{\partial^2 \bar u}{\partial \bar{z}^2} = \bar{f}.
\end{align}
We set the force to vanish, $\bar{f} = 0$, as we are interested in the static deformation and the eigenmodes of vibrations around it.

Let us now assume $\bar{u}(\bar z, \bar t) = \bar{u}_0(\bar z) + \bar{u}_1(\bar z, \bar t)$ where $\bar{u}_0 \gg \bar{u}_1$ as in the main text. 
Using this assumption, we may insert the relation for $\bar{u}(\bar z, \bar t)$ to Eq.~\eqref{eq:dimensionlessEB} and separate the resulting equation into two by considering the zeroth and first power of $\bar{u}_1$.
We find that the static deformation $\bar u_0$ is determined by the zeroth order equation which reads 
\begin{equation} \label{eq:app:u0eq}
 \partial_{\bar{z}\bar{z}\bar{z}\bar{z}} \bar{u}_0 + \bar{\alpha}^2\, \partial_{\bar{z}\bar{z}} \bar{u}_0=0,\quad
 \bar{\alpha}^2= \bar{P} - \frac{1}{2} \int_0^1 d\bar{z} \left( \partial_{\bar{z}} \bar{u}_0 \right)^2.
\end{equation}
A static deformation is possible only if $\bar{\alpha}^2 > 0$; otherwise $\bar {u}_0 = 0$ is the only solution.
This is known as the buckling transition.

We may begin solving Eq.~\eqref{eq:app:u0eq} by assuming that $\bar{\alpha}^2$ is independent of $u_0$.
If indeed $\bar{\alpha}^2 > 0$, one can provide a general solution of the differential equation in terms of trigonometric and linear functions.
However, the boundary conditions~\eqref{eq:non-linear_boundary_cond} fixes $\bar{\alpha} = 2 \pi$ for a single anti-node solution which reads
\begin{equation} 
    \bar{u}_0(\bar{z}) = \bar{u}_m\frac{1 - \cos(\bar{\alpha} \bar{z})}{2}.
\end{equation}
Here, the parameter $\bar{u}_m = \pm 4\sqrt{\frac{\bar{P}}{\bar{\alpha}^2} -1}$ is fixed self-consistently by inserting the solution to the definition of $\bar{\alpha}^2$.
It also describes the value $\bar{u}_0$ of deformation in the middle of the beam $\bar{z} = 1/2$.
Note that $\bar{P}$ may be removed in favour of $\bar{u}_m$ if the condition for buckling $\bar{P} > \bar{\alpha}^2 = 4\pi^2$ holds true.

In the first order of $\bar{u}_1$, its dynamics is described by
\begin{subequations}
\begin{eqnarray}
 &&\partial_{\bar{t}\bar{t}} u_1 + \partial_{\bar{z}\bar{z}\bar{z}\bar{z}} \bar{u}_1 +\bar{\alpha}^2\, \partial_{\bar{z}\bar{z}} \bar{u}_1
 -\bar{\beta}\, \partial_{\bar{z}\bar{z}} \bar{u}_0 = 0,\\
 && \bar{\beta}=\int_0^1 d\bar{z} \left( \partial_{\bar{z}} \bar{u}_0\right) \left( \partial_{\bar{z}} \bar{u}_1 \right).
 \label{eq:self-consistency_beta}
\end{eqnarray}
\end{subequations}
We can now find the flexural eigenmodes by Fourier transforming $\bar{u}_1(\bar{z},\bar{t})=\bar{u}_1(\bar{z})\, e^{-i\, \bar{\omega}\, \bar{t}}$
which gives
\begin{equation}
 \partial_{\bar{z}\bar{z}\bar{z}\bar{z}} \bar{u}_1 + \bar{\alpha}^2 \partial_{\bar{z}\bar{z}} \bar{u}_1 - \bar{\omega}^2\, \bar{u}_1
 =  \bar{\beta}\, \partial_{\bar{z}\bar{z}} \bar{u}_0.
 \label{eqn:eigenmode_equation}
\end{equation}
Here the dimensionless frequency $\bar{\omega}$ is related to the physical one with $\omega = \sqrt{\frac{E\, I_x}{\rho\, A\, L^4}} \bar{\omega}$.
The differential equation~\eqref{eqn:eigenmode_equation} is written so that the left hand side is the homogeneous equation while the right hand side provides the non-homogeneous term.
It is straightforward to check that setting $\bar{u}_1 \propto \partial_{\bar{z}\bar{z}} \bar{u}_0$ gives a particular solution.
Thus, the general solution may be written as
\begin{align}
 \bar{u}_1(\bar{z}) &= C_1 \cos (\delta_+  \bar{z})+C_2 \sin (\delta_+ \bar{z}) + C_3 \cosh (\delta_-  \bar{z}) \notag\\ 
 &\quad+ C_4 \sinh (\delta_-  \bar{z})
  + C_5 \frac{\bar{\alpha}^2\bar{u}_m}{2} \cos{\bar{\alpha}\bar{z}},
 \label{eqn:general_eigenmodes}
\end{align}
where 
\begin{equation}
    \delta_\pm = \frac{\sqrt{\sqrt{\bar{\alpha}^4+4 \bar{\omega} ^2} \pm \bar{\alpha} ^2}}{\sqrt{2}}.
\end{equation}
Here, the terms with $C_1 \dots C_4$ specify the solution of the homogeneous equation.

The constants $\bm{C}=(C_1,C_2,C_3,C_4,C_5)$ may be fixed in the following manner:
The boundary conditions for $\bar{u}_1$ give in total four equations.
The last equation is obtained by inserting the full general solution into Eq.~\eqref{eqn:eigenmode_equation} where we use the definition~\eqref{eq:self-consistency_beta} for $\bar{\beta}$.
This set of equations can be rearranged into a linear equation $\bm{M}_{\text{nl}}\bm{C}=\bm{0}$, where the matrix $\bm{M}_{\text{nl}}$ depends on frequency $\bar\omega$.
One can then solve the equation numerically in many ways: we use the singular value decomposition to minimize the smallest singular value and use the corresponding vector $\bm{C}$ belonging into the null space of $\bm{M}_{\text{nl}}$.
We then normalize the found solution for $\bar{u}_1$ so that it corresponds to the eigenmode $\chi_n$, that is, $\int_0^1 d \bar{z} \chi_n^2 = 1$.

We note that it would be possible to remove $C_5$ by setting $C_5 = -\bar{\beta}/\bar{\omega}^2$ which gives the correct particular solution and, then, self-consistently solve for $\bar{\beta}$.
However, the division by the unknown frequency is numerically problematic since values $\bar{\omega} \approx 0$ are needed.
Even with finite frequencies, this method did not produce orthogonal modes.
With the method described above, we find the eigenmodes $\chi_n$ to be orthonormal to a reasonable numerical accuracy, meaning that $\int_0^1 d \bar{z} \chi_n \chi_m \approx \delta_{nm}$.

\begin{figure}
    \centering
    \includegraphics{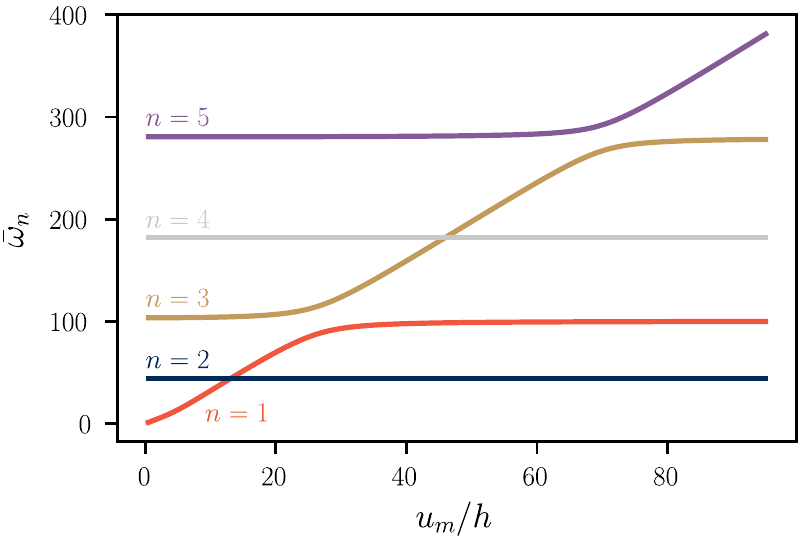}
    \caption{The dimensionless eigenfrequencies of the dynamical mode $\bar{u}_1$.}
    \label{fig:mech_eigenfr}
\end{figure}

In Fig.~\ref{fig:mech_eigenfr} we have plotted the five first eigenfrequencies.
The structure of eigenfrequencies as a function of the static deformation $u_m$ is that of crossings and avoided crossings.
We find, as in Ref.~\onlinecite{nayfeh2008exact}, that the antisymmetric modes with $n = 2, 4 \dots$ do not depend on the static deformation.
This is because the static deformation is symmetric and, thus, $\bar \beta = 0$ for these modes.
As $u_m$ increases, there is a crossing in eigenfrequencies between the symmetric and antisymmetric mode, followed by an avoided crossing with the next symmetric mode.

\begin{figure}
    \centering
    \includegraphics{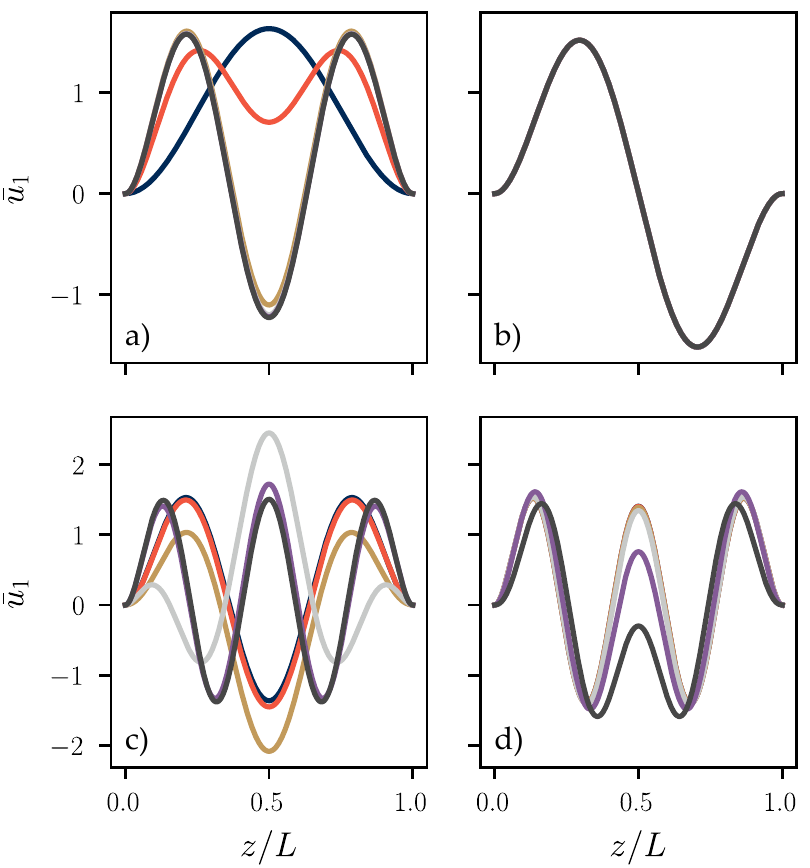}
    \caption{Eigenmodes $\chi_n$ for a) $n = 1$, b) $n = 2$, c) $n = 3$, and d) $n = 5$.
    Here, $u_m/h$ are chosen from approximately $\{0.5, 19, 37, 55, 73, 91 \}$.}
    \label{fig:mech_eigenmodes}
\end{figure}

A few examples of the eigenmodes $\chi_n$ for different static deformations $u_m$ are plotted in Fig.~\ref{fig:mech_eigenmodes}.
In general, we note that the eigenmodes $\chi_n$ with odd $n$ are changed by the static deformation while $\chi_n$ with even $n$ remain the same.
More specifically, the first mode $\chi_1$ is plotted in Fig.~\ref{fig:mech_eigenmodes}a). 
After the crossing in eigenfrequencies $\omega_1$ and $\omega_2$ around $u_m/h \approx 14$ the eigenmode $\chi_1$ starts to resemble the third mode at small $u_m$: it has three anti-nodes. 
The same can be observed for $\chi_3$ in Fig.~\ref{fig:mech_eigenmodes}c) where the eigenmodes have either three or five anti-nodes.
On the other hand, the eigenmode $\chi_2$ remains independent of $u_m/h$ as can be seen in Fig.~\ref{fig:mech_eigenmodes}b).

\section{Magnetic dynamics}

\subsection{Magnetic hysteresis}
\label{sec:magnetic_hysteresis}

The static component of the magnetization $\bm{M}_0$ can be determined for a given external field $\bm{H}_0$, knowing the saturation magnetization $M_S$, the magnetoelastic constant $\Bc$ and the static strain $\bar{\epsilon}_{zz}^{(0)}$.

For simplicity, we assume that the static component of the magnetic
field is parallel to the $yz$-plane: $\bm{H}_0=H_0 (\cos\phi\,
\hat{z} + \sin\phi\, \hat{y})$ with $H_0>0$. In this case, the static magnetization $\bm{M}_0$ is obtained by minimizing the magnetic free energy, proportional to Eq.~\eqref{eq:averagemagneticF} with $\bm{H} = \bm{H}_0$ and $\bm{M} = \bm{M}_0$, which is up to a constant
\begin{eqnarray}
    \mathcal{F}_0 =&& \, -\mu_0 \bm{H}_0\cdot \bm{M}_0 + \mu_0\left(\frac{1}{2}- \bar{\epsilon}_{zz}^{(0)}\right) M_x^2 \nonumber\\
    &&\, + \mu_0\left(\frac{\Bc}{\mu_0 M_S^2} + 1\right) \bar{\epsilon}_{zz}^{(0)} M_z^2.
\end{eqnarray}
In the validity range of the Euler-Bernoulli theory $\partial_z u(z)\ll 1$ and $\bar{\epsilon}_{zz}^{(0)}\ll 1$, we
must have $M_{0x}=0$.
Thus, we can make the ansatz $\bm{M}_0=M_S(\cos\theta\, \hat{z}+
\sin\theta\, \hat{y})$ where the angle $\theta$ is obtained by
minimizing
\begin{equation}
    \frac{\mathcal{F}_0}{\mu_0 M_S H_0} =
    - \cos(\theta-\phi) + 
     \left(\frac{\Bc}{\mu_0 M_S^2} + 1\right) \frac{M_S}{H_0} \bar{\epsilon}_{zz}^{(0)} \cos^2\theta.
    \label{eq:app:magF0}
\end{equation}      
Since, in general, the magnetoelastic constant $B_1$ can be positive or negative, the solution of $\theta$ depends on this choice.
More precisely, the sign of $B_1/(\mu_0 M_S^2) + 1$ determines the behavior of the magnetic free energy $\mathcal{F}_0$ when $H_0$ and $\bar{\epsilon}_{zz}^{(0)}$ are fixed and non-zero.
In either case, the competition between the external magnetic field, the demagnetizing field, and the magnetoelastic field gives rise to hysteresis that is similar to the Stoner–Wohlfarth hysteresis~\cite{stoner1948mechanism, wohlfarth1958relations}.

\begin{figure}
    \centering
    \includegraphics{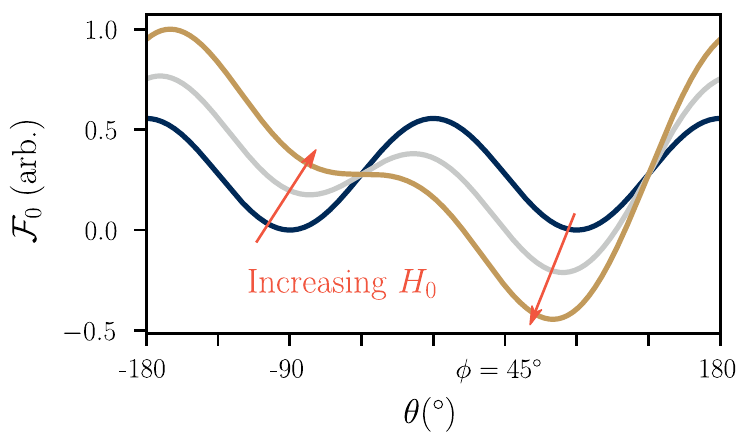}
    \caption{The static magnetic free energy derived in Eq.~\eqref{eq:app:magF0} for $B_1 > -\mu_0 M_S^2$ (corresponding to a hard $z$-axis) as a function of the angle $\theta$, as the magnitude $H_0$ of the external field pointing at $\phi = 45^\circ$ is increased (orange arrows). The blue line corresponds to $H_0 = 0$.
    As $H_0$ is increased, the metastable local minimum of $\mathcal{F}_0$ disappears and the global minimum approaches~$\phi$.}
    \label{fig:stat_m_energy}
\end{figure}

If $B_1 > - \mu_0 M_S^2$, the prefactor of $\cos^2\theta$ term in Eq.~\eqref{eq:app:magF0} is positive and, therefore, the $z$ axis is a hard axis of the magnetic system.
The magnetic free energy in this situation is depicted in Fig.~\ref{fig:stat_m_energy} as a function of $\theta$.
Without any external field (blue line), the minima are found at $\theta = \pm 90^\circ$.
When the external field magnitude $H_0$ is increased, at first, one of the minima becomes a global minimum and, eventually, the second metastable local minimum disappears.
At the same time, the magnetization angle $\theta$ corresponding to a global minimum approaches $\phi = 45^\circ$.

The case with $B_1 < - \mu_0 M_S^2$ corresponds to an easy $z$~axis, as the sign of $B_1/(\mu_0 M_S^2)+1$ is negative.
Mathematically, this case can be mapped exactly to the previous hard $z$~axis case with the transformations $\theta \rightarrow \theta + 180^\circ$ and $\mathcal{F}_0 \rightarrow - \mathcal{F}_0$.
Thus, comparing to Fig.~\ref{fig:stat_m_energy}, the maxima correspond to the minima after a shift in $\theta$ in this case.

For $\phi=90^\circ$, i.e., magnetic field perpendicular to the beam, the problem can be solved analytically. For low
  fields, the (meta)stable magnetization directions depend on the sign
  of the anisotropy term, i.e., the sign of
  $B_1/(\mu_0 M_S^2) +1$, because when this term is positive, the beam axis
  is a hard axis and the magnetization prefers to lie along the
  magnetic field. For the opposite sign of the anisotropy, the static
  magnetization is along the beam for low fields and along the field
  for high fields. The coercive field, i.e., the size of the magnetic field where the second relative
  minimum disappear is however in both cases
  \begin{equation}
    H_{c} = 2 \left\lvert \frac{B_1}{\mu_0 M_S^2}+1 \right\rvert \bar{\epsilon}_{zz}^{(0)} M_S.
    \label{eq:coercivefield}
  \end{equation}
  This field can be accessed in magnetic hysteresis measurements, but
  it is also related with the size of the magnomechanical coupling.

\begin{figure}
    \centering
    \includegraphics{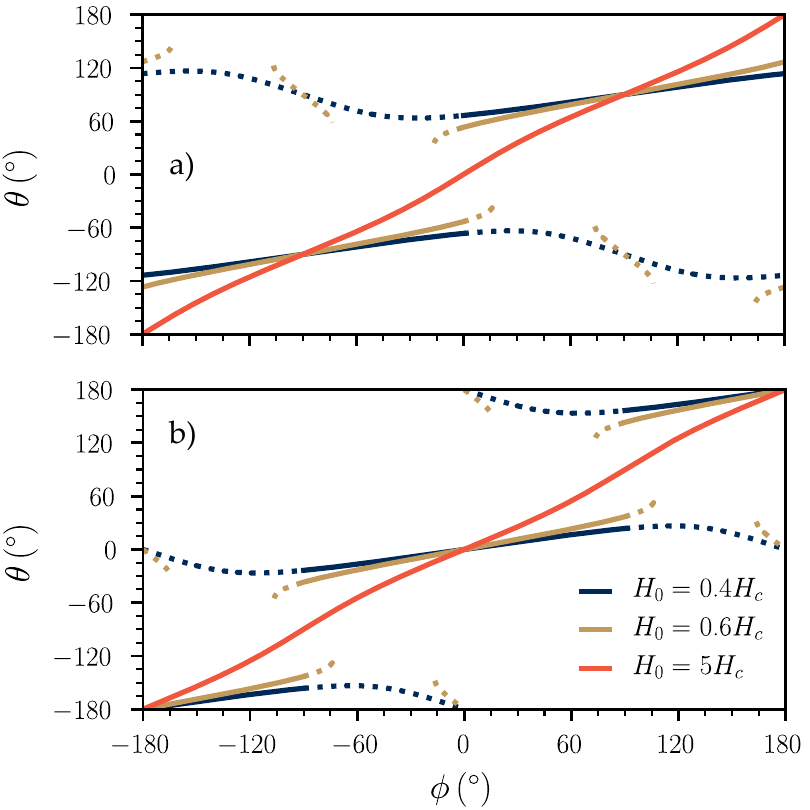}
    \caption{The magnetization angle $\theta$ as a function of $\phi$. In a) $B_1 < -\mu_0 M_S^2$ and in b) $B_1 > -\mu_0 M_S^2$ corresponding to the $z$~axis being a hard and easy axis, respectively. The dotted lines represent the metastable free energy minimum. For $H_0 > H_c$, the metastable minimum disappears, and for large $H_0$ we find $\theta \approx \phi$ as expected.}
    \label{fig:thetavsphi}
\end{figure}

In Fig.~\ref{fig:thetavsphi} we show the solution of $\theta$ as a function of $\phi$ for different external field magnitudes $H_0$.
The magnetization angle~$\theta$ is fully characterized by $\phi$ and $H_0/H_c$ if the $z$~axis being a hard or easy axis is given (the sign of $B_1/(\mu_0 M_S^2) + 1$).

\subsection{Magnetic Hamiltonian}

\label{sec:magnetic_hamiltonian}

The magnetization dynamics is given by the Landau-Lifshitz equation that can be written in the Hamiltonian formalism~\cite{stancil2009spin}
\begin{equation}
\label{eq:class_magn_dynam}
    \dot{a}=i\, \frac{\partial H_{mg}}{\partial a^\ast},\qquad
    \dot{a}^\ast=-i\, \frac{\partial H_{mg}}{\partial a},
\end{equation}
where
\begin{subequations}
\label{eq:LL-Hamiltonia_rel}
\begin{eqnarray}
   && M_{z'}= M_S-\frac{\gammah}{L A} a^\ast a,\\
   && M_{+'}=\sqrt{\frac{2\gammah M_S}{L A}} \sqrt{1-\frac{\gammah}{2 M_S L A} a^\ast a}\, a,\\
   && M_{-'}=\sqrt{\frac{2 \gammah M_S}{L A}} a^\ast \sqrt{1-\frac{\gammah}{2 M_S L A} a^\ast a},\\
   && H_{mg}=L A \mathcal{F},\quad
   M_{\pm'}=M_{x'}\pm i\, M_{y'}
\end{eqnarray}
\end{subequations}
and the conjugate variables are $a=(q+i p)/\sqrt{2}$, $a=(q-i p)/\sqrt{2}$. The quantization can be achieved with $[\hat{a},\hat{a}^\dagger]=i\, \hbar\, \{a,a^\ast\}=\hbar$, where the braces represent the Poisson brackets. This corresponds to the Holstein-Primakoff bosonization, where $\hat{m}^\dagger=:\hat{a}^\dagger/\sqrt{\hbar}$ creates a boson.

If we assume $\bm{M}_0=M_S\, \hat{z}'$ and  $H_0\gg h$, we can obtain the linearized Landau-Lifshitz equation by retaining only the quadratic terms in $a,\, a^\ast$ in Eq.~\eqref{eq:LL-Hamiltonia_rel}. After a linear transformation $a \to \tilde{a}$, having the same form as Eq.~\eqref{eq:diagonal_transfor}, the Hamiltonian reads $H_{mg}=\omega_K \tilde{a}^\ast \tilde{a} -(\tilde{h}^\ast \tilde{a} + \tilde{h} \tilde{a}^\ast)$, where
\begin{equation}
    \tilde{h}=
    \mu_0 \sqrt{\frac{\gammah M_S L A}{2}} \left[ h_{x'} \left( \zeta_+-\zeta_- \right)
    +i h_{y'} \left( \zeta_+ + \zeta_- \right) \right]
\end{equation}
and $\omega_K,\, \zeta_\pm$ are defined in the main text in Eqs.~\eqref{eq:Kittel_freq} and~\eqref{eq:diagonal_transfor}. Then, by Fourier transforming $\tilde{a}\to e^{-i \omega t} \tilde{a}$, $h_{x',y'}\to e^{-i \omega t} h_{x',y'}$, the dynamic equations for the magnetization~\eqref{eq:class_magn_dynam} are
\begin{equation}
    \begin{pmatrix}
    \tilde{h}^\ast\\
    \tilde{h}
    \end{pmatrix}
 =
 \begin{pmatrix}
 \omega_K-\omega  &&  0\\
 0                &&  \omega_K-\omega
 \end{pmatrix}
 \begin{pmatrix}
 \tilde{a}^\ast\\
 \tilde{a}
 \end{pmatrix}
\end{equation}
from which we can see that $\omega_K$ is the FMR resonance frequency.

\section{Derivation of the linear response}
\label{sec:inputoutputdetails}

As in the main text, let us focus on the case of a single flexural mode.
Using the input-output equations given in Eqs.~\eqref{eq:io:magnon} and~\eqref{eq:io:vibrations} with the Hamiltonian~\eqref{eq:linearizedHmm}, we obtain a system of equations
\begin{align}
    \begin{pmatrix}
        \dot{\hat{x}}_l \\ \dot{\hat{p}}_l \\ \dot{\hat{x}}_b \\ \dot{\hat{p}}_b
    \end{pmatrix}
    = M \begin{pmatrix}
         \hat{x}_l \\  \hat{p}_l \\ \hat{x}_b \\  \hat{p}_b
    \end{pmatrix}
    -
    C \begin{pmatrix}
         \hat{x}_{l,\mathrm{in}} \\  \hat{p}_{l,\mathrm{in}} \\ \hat{x}_{b,\mathrm{in}} \\  \hat{p}_{b,\mathrm{in}}
    \end{pmatrix},
    \label{eq:matrixEOMtime}
\end{align}
where $C = \mathrm{diag}(\sqrt{\kappa_e},\sqrt{\kappa_e},\sqrt{\gamma_e},\sqrt{\gamma_e})$ is a diagonal matrix containing the external coupling rates and
\begin{align}
    M = \begin{pmatrix}
        -\kappa/2 & \Delta & 0 & 0 \\
         -\Delta & -\kappa/2 & G & 0 \\
        0 & 0 & -\gamma/2 & \omega_m \\
        G & 0 & -\omega_m & -\gamma/2 \\
    \end{pmatrix},
\end{align}
where $\Delta = \omega_K - \omega_d$.
The quadrature operators are connected to their bosonic counterparts by a unitary transformation.
Let us now denote this transformation by $U$ and define it such that
\begin{align}
    \begin{pmatrix}
         \hat{x}_l \\  \hat{p}_l \\ \hat{x}_b \\  \hat{p}_b
    \end{pmatrix}
    = U \begin{pmatrix}
         \hat{l} \\  \hat{l}^\dagger \\ \hat{b} \\  \hat{b}^\dagger
    \end{pmatrix},\,
    U = \frac{1}{\sqrt{2}}\begin{pmatrix}
        1 & 1 & 0 & 0 \\
         -i & i & 0 & 0 \\
        0 & 0 &  1 & 1 \\
        0 & 0 & -i & i \\
        \end{pmatrix}.
        \label{eq:quadraturetransformation}
\end{align}
The same relation holds for the input operators.
Now, we can apply the Fourier transformation $\hat{x}^{(\dagger)}(\omega) = \int \hat{x}^{(\dagger)}(t) e^{i \omega t} dt$
to Eq.~\eqref{eq:matrixEOMtime} and use the transformation~\eqref{eq:quadraturetransformation} to express the linear response matrix $T$ as
\begin{align}
    T(\omega) = U^{-1} (i \omega I + M)^{-1} C U,
\end{align}
where $I$ is the $4 \times 4$ identity matrix.
Due to the complex structure of the response matrix $T$, there are only six independent components which read
\begin{subequations}
\begin{align}
    T_{11} &= -\frac{i\sqrt{\kappa_e}}{D} \left[\frac{\omega_m}{2} G^2 + i\chi_\gamma(-\omega_m)\chi_\gamma(\omega_m)\chi_\kappa(-\Delta)\right],\\
    T_{12} &= -\frac{i\sqrt{\kappa_e}}{D} \frac{\omega_m}{2} G^2,\\
    T_{13} &= -\frac{i\sqrt{\gamma_e}}{D} \frac{G}{2} \chi_\gamma(-\omega_m)\chi_\kappa(-\Delta), \\
    T_{14} &= -\frac{i\sqrt{\gamma_e}}{D} \frac{G}{2} \chi_\gamma(\omega_m)\chi_\kappa(-\Delta), \\
    T_{33} &= -\frac{i\sqrt{\gamma_e}}{D} \left[\frac{\Delta}{2} G^2 + i\chi_\gamma(-\omega_m)\chi_\kappa(\Delta)\chi_\kappa(-\Delta)\right],\\
    T_{34} &= -\frac{i\sqrt{\gamma_e}}{D} \frac{\Delta}{2} G^2,
\end{align}
\end{subequations}
where $\chi_\eta(\omega_0) = i(\omega - \omega_0) - \eta/2$ and 
\begin{align}
    D = \chi_\gamma(\omega_m)\chi_\gamma(-\omega_m)\chi_\kappa(\Delta)\chi_\kappa(-\Delta) - \Delta \omega_m G^2
\end{align}
is the determinant of the matrix $i\omega I + M$.
The full linear response matrix $T$ may now be written as
\begin{align}
    T = \begin{pmatrix}
         T_{11}(\omega) & T_{12}(\omega) & T_{13}(\omega) & T_{14}(\omega) \\
         T_{12}^*(\omega) & T_{11}^*(-\omega) & T_{14}^*(\omega) & T_{13}^*(-\omega) \\
         \sqrt{\frac{\kappa_e}{\gamma_e}}T_{13}(\omega) & \sqrt{\frac{\kappa_e}{\gamma_e}}T_{14}(\omega) & T_{33}(\omega) & T_{34}(\omega) \\
         \sqrt{\frac{\kappa_e}{\gamma_e}}T_{14}^*(\omega) & \sqrt{\frac{\kappa_e}{\gamma_e}}T_{13}^*(-\omega) & T_{34}^*(\omega) & T_{33}^*(-\omega)
    \end{pmatrix}.
\end{align}
That is, the other elements are obtained by conjugation, multiplying by a prefactor $\sqrt{\frac{\kappa_e}{\gamma_e}}$, and setting $\omega \rightarrow - \omega$.


\begin{thebibliography}{53}%
\makeatletter
\providecommand \@ifxundefined [1]{%
 \@ifx{#1\undefined}
}%
\providecommand \@ifnum [1]{%
 \ifnum #1\expandafter \@firstoftwo
 \else \expandafter \@secondoftwo
 \fi
}%
\providecommand \@ifx [1]{%
 \ifx #1\expandafter \@firstoftwo
 \else \expandafter \@secondoftwo
 \fi
}%
\providecommand \natexlab [1]{#1}%
\providecommand \enquote  [1]{``#1''}%
\providecommand \bibnamefont  [1]{#1}%
\providecommand \bibfnamefont [1]{#1}%
\providecommand \citenamefont [1]{#1}%
\providecommand \href@noop [0]{\@secondoftwo}%
\providecommand \href [0]{\begingroup \@sanitize@url \@href}%
\providecommand \@href[1]{\@@startlink{#1}\@@href}%
\providecommand \@@href[1]{\endgroup#1\@@endlink}%
\providecommand \@sanitize@url [0]{\catcode `\\12\catcode `\$12\catcode
  `\&12\catcode `\#12\catcode `\^12\catcode `\_12\catcode `\%12\relax}%
\providecommand \@@startlink[1]{}%
\providecommand \@@endlink[0]{}%
\providecommand \url  [0]{\begingroup\@sanitize@url \@url }%
\providecommand \@url [1]{\endgroup\@href {#1}{\urlprefix }}%
\providecommand \urlprefix  [0]{URL }%
\providecommand \Eprint [0]{\href }%
\providecommand \doibase [0]{https://doi.org/}%
\providecommand \selectlanguage [0]{\@gobble}%
\providecommand \bibinfo  [0]{\@secondoftwo}%
\providecommand \bibfield  [0]{\@secondoftwo}%
\providecommand \translation [1]{[#1]}%
\providecommand \BibitemOpen [0]{}%
\providecommand \bibitemStop [0]{}%
\providecommand \bibitemNoStop [0]{.\EOS\space}%
\providecommand \EOS [0]{\spacefactor3000\relax}%
\providecommand \BibitemShut  [1]{\csname bibitem#1\endcsname}%
\let\auto@bib@innerbib\@empty
\bibitem [{\citenamefont {Aspelmeyer}\ \emph {et~al.}(2014)\citenamefont
  {Aspelmeyer}, \citenamefont {Kippenberg},\ and\ \citenamefont
  {Marquardt}}]{aspelmeyer2014cavity}%
  \BibitemOpen
  \bibfield  {author} {\bibinfo {author} {\bibfnamefont {M.}~\bibnamefont
  {Aspelmeyer}}, \bibinfo {author} {\bibfnamefont {T.~J.}\ \bibnamefont
  {Kippenberg}},\ and\ \bibinfo {author} {\bibfnamefont {F.}~\bibnamefont
  {Marquardt}},\ }\bibfield  {title} {\bibinfo {title} {Cavity optomechanics},\
  }\href@noop {} {\bibfield  {journal} {\bibinfo  {journal} {Rev. Mod. Phys.}\
  }\textbf {\bibinfo {volume} {86}},\ \bibinfo {pages} {1391} (\bibinfo {year}
  {2014})}\BibitemShut {NoStop}%
\bibitem [{\citenamefont {Teufel}\ \emph {et~al.}(2011)\citenamefont {Teufel},
  \citenamefont {Donner}, \citenamefont {Li}, \citenamefont {Harlow},
  \citenamefont {Allman}, \citenamefont {Cicak}, \citenamefont {Sirois},
  \citenamefont {Whittaker}, \citenamefont {Lehnert},\ and\ \citenamefont
  {Simmonds}}]{Teufel2011b}%
  \BibitemOpen
  \bibfield  {author} {\bibinfo {author} {\bibfnamefont {J.~D.}\ \bibnamefont
  {Teufel}}, \bibinfo {author} {\bibfnamefont {T.}~\bibnamefont {Donner}},
  \bibinfo {author} {\bibfnamefont {D.}~\bibnamefont {Li}}, \bibinfo {author}
  {\bibfnamefont {J.~W.}\ \bibnamefont {Harlow}}, \bibinfo {author}
  {\bibfnamefont {M.~S.}\ \bibnamefont {Allman}}, \bibinfo {author}
  {\bibfnamefont {K.}~\bibnamefont {Cicak}}, \bibinfo {author} {\bibfnamefont
  {A.~J.}\ \bibnamefont {Sirois}}, \bibinfo {author} {\bibfnamefont {J.~D.}\
  \bibnamefont {Whittaker}}, \bibinfo {author} {\bibfnamefont {K.~W.}\
  \bibnamefont {Lehnert}},\ and\ \bibinfo {author} {\bibfnamefont {R.~W.}\
  \bibnamefont {Simmonds}},\ }\bibfield  {title} {\bibinfo {title} {{Sideband
  cooling of micromechanical motion to the quantum ground state}},\ }\href@noop
  {} {\bibfield  {journal} {\bibinfo  {journal} {Nature}\ }\textbf {\bibinfo
  {volume} {475}},\ \bibinfo {pages} {359} (\bibinfo {year}
  {2011})}\BibitemShut {NoStop}%
\bibitem [{\citenamefont {Chan}\ \emph {et~al.}(2011)\citenamefont {Chan},
  \citenamefont {Alegre}, \citenamefont {Safavi-Naeini}, \citenamefont {Hill},
  \citenamefont {Krause}, \citenamefont {Gr\"oblacher}, \citenamefont
  {Aspelmeyer},\ and\ \citenamefont {Painter}}]{AspelmeyerCool11}%
  \BibitemOpen
  \bibfield  {author} {\bibinfo {author} {\bibfnamefont {J.}~\bibnamefont
  {Chan}}, \bibinfo {author} {\bibfnamefont {T.~P.~M.}\ \bibnamefont {Alegre}},
  \bibinfo {author} {\bibfnamefont {A.~H.}\ \bibnamefont {Safavi-Naeini}},
  \bibinfo {author} {\bibfnamefont {J.~T.}\ \bibnamefont {Hill}}, \bibinfo
  {author} {\bibfnamefont {A.}~\bibnamefont {Krause}}, \bibinfo {author}
  {\bibfnamefont {S.}~\bibnamefont {Gr\"oblacher}}, \bibinfo {author}
  {\bibfnamefont {M.}~\bibnamefont {Aspelmeyer}},\ and\ \bibinfo {author}
  {\bibfnamefont {O.}~\bibnamefont {Painter}},\ }\bibfield  {title} {\bibinfo
  {title} {Laser cooling of a nanomechanical oscillator into its quantum ground
  state},\ }\href@noop {} {\bibfield  {journal} {\bibinfo  {journal} {Nature}\
  }\textbf {\bibinfo {volume} {478}} (\bibinfo {year} {2011})}\BibitemShut
  {NoStop}%
\bibitem [{\citenamefont {Ockeloen-Korppi}\ \emph {et~al.}(2018)\citenamefont
  {Ockeloen-Korppi}, \citenamefont {Damsk{\"a}gg}, \citenamefont
  {Pirkkalainen}, \citenamefont {Asjad}, \citenamefont {Clerk}, \citenamefont
  {Massel}, \citenamefont {Woolley},\ and\ \citenamefont
  {Sillanp{\"a}{\"a}}}]{Entanglement}%
  \BibitemOpen
  \bibfield  {author} {\bibinfo {author} {\bibfnamefont {C.~F.}\ \bibnamefont
  {Ockeloen-Korppi}}, \bibinfo {author} {\bibfnamefont {E.}~\bibnamefont
  {Damsk{\"a}gg}}, \bibinfo {author} {\bibfnamefont {J.~M.}\ \bibnamefont
  {Pirkkalainen}}, \bibinfo {author} {\bibfnamefont {M.}~\bibnamefont {Asjad}},
  \bibinfo {author} {\bibfnamefont {A.~A.}\ \bibnamefont {Clerk}}, \bibinfo
  {author} {\bibfnamefont {F.}~\bibnamefont {Massel}}, \bibinfo {author}
  {\bibfnamefont {M.~J.}\ \bibnamefont {Woolley}},\ and\ \bibinfo {author}
  {\bibfnamefont {M.~A.}\ \bibnamefont {Sillanp{\"a}{\"a}}},\ }\bibfield
  {title} {\bibinfo {title} {Stabilized entanglement of massive mechanical
  oscillators},\ }\href@noop {} {\bibfield  {journal} {\bibinfo  {journal}
  {Nature}\ }\textbf {\bibinfo {volume} {556}},\ \bibinfo {pages} {478}
  (\bibinfo {year} {2018})}\BibitemShut {NoStop}%
\bibitem [{\citenamefont {Riedinger}\ \emph {et~al.}(2018)\citenamefont
  {Riedinger}, \citenamefont {Wallucks}, \citenamefont {Marinkovi{\'c}},
  \citenamefont {L{\"o}schnauer}, \citenamefont {Aspelmeyer}, \citenamefont
  {Hong},\ and\ \citenamefont {Gr{\"o}blacher}}]{Groblacher}%
  \BibitemOpen
  \bibfield  {author} {\bibinfo {author} {\bibfnamefont {R.}~\bibnamefont
  {Riedinger}}, \bibinfo {author} {\bibfnamefont {A.}~\bibnamefont {Wallucks}},
  \bibinfo {author} {\bibfnamefont {I.}~\bibnamefont {Marinkovi{\'c}}},
  \bibinfo {author} {\bibfnamefont {C.}~\bibnamefont {L{\"o}schnauer}},
  \bibinfo {author} {\bibfnamefont {M.}~\bibnamefont {Aspelmeyer}}, \bibinfo
  {author} {\bibfnamefont {S.}~\bibnamefont {Hong}},\ and\ \bibinfo {author}
  {\bibfnamefont {S.}~\bibnamefont {Gr{\"o}blacher}},\ }\bibfield  {title}
  {\bibinfo {title} {Remote quantum entanglement between two micromechanical
  oscillators},\ }\href@noop {} {\bibfield  {journal} {\bibinfo  {journal}
  {Nature}\ }\textbf {\bibinfo {volume} {556}},\ \bibinfo {pages} {473}
  (\bibinfo {year} {2018})}\BibitemShut {NoStop}%
\bibitem [{\citenamefont {Mirhosseini}\ \emph {et~al.}(2020)\citenamefont
  {Mirhosseini}, \citenamefont {Sipahigil}, \citenamefont {Kalaee},\ and\
  \citenamefont {Painter}}]{Painter2020scqb}%
  \BibitemOpen
  \bibfield  {author} {\bibinfo {author} {\bibfnamefont {M.}~\bibnamefont
  {Mirhosseini}}, \bibinfo {author} {\bibfnamefont {A.}~\bibnamefont
  {Sipahigil}}, \bibinfo {author} {\bibfnamefont {M.}~\bibnamefont {Kalaee}},\
  and\ \bibinfo {author} {\bibfnamefont {O.}~\bibnamefont {Painter}},\
  }\bibfield  {title} {\bibinfo {title} {Superconducting qubit to optical
  photon transduction},\ }\href@noop {} {\bibfield  {journal} {\bibinfo
  {journal} {Nature}\ }\textbf {\bibinfo {volume} {588}},\ \bibinfo {pages}
  {599} (\bibinfo {year} {2020})}\BibitemShut {NoStop}%
\bibitem [{\citenamefont {Lecocq}\ \emph {et~al.}(2016)\citenamefont {Lecocq},
  \citenamefont {Clark}, \citenamefont {Simmonds}, \citenamefont {Aumentado},\
  and\ \citenamefont {Teufel}}]{Teufel2016fConv}%
  \BibitemOpen
  \bibfield  {author} {\bibinfo {author} {\bibfnamefont {F.}~\bibnamefont
  {Lecocq}}, \bibinfo {author} {\bibfnamefont {J.~B.}\ \bibnamefont {Clark}},
  \bibinfo {author} {\bibfnamefont {R.~W.}\ \bibnamefont {Simmonds}}, \bibinfo
  {author} {\bibfnamefont {J.}~\bibnamefont {Aumentado}},\ and\ \bibinfo
  {author} {\bibfnamefont {J.~D.}\ \bibnamefont {Teufel}},\ }\bibfield  {title}
  {\bibinfo {title} {Mechanically mediated microwave frequency conversion in
  the quantum regime},\ }\href@noop {} {\bibfield  {journal} {\bibinfo
  {journal} {Phys. Rev. Lett.}\ }\textbf {\bibinfo {volume} {116}},\ \bibinfo
  {pages} {043601} (\bibinfo {year} {2016})}\BibitemShut {NoStop}%
\bibitem [{\citenamefont {Ockeloen-Korppi}\ \emph {et~al.}(2016)\citenamefont
  {Ockeloen-Korppi}, \citenamefont {Damsk\"agg}, \citenamefont {Pirkkalainen},
  \citenamefont {Heikkil\"a}, \citenamefont {Massel},\ and\ \citenamefont
  {Sillanp\"a\"a}}]{CasparAmp}%
  \BibitemOpen
  \bibfield  {author} {\bibinfo {author} {\bibfnamefont {C.~F.}\ \bibnamefont
  {Ockeloen-Korppi}}, \bibinfo {author} {\bibfnamefont {E.}~\bibnamefont
  {Damsk\"agg}}, \bibinfo {author} {\bibfnamefont {J.-M.}\ \bibnamefont
  {Pirkkalainen}}, \bibinfo {author} {\bibfnamefont {T.~T.}\ \bibnamefont
  {Heikkil\"a}}, \bibinfo {author} {\bibfnamefont {F.}~\bibnamefont {Massel}},\
  and\ \bibinfo {author} {\bibfnamefont {M.~A.}\ \bibnamefont
  {Sillanp\"a\"a}},\ }\bibfield  {title} {\bibinfo {title} {Low-noise
  amplification and frequency conversion with a multiport microwave
  optomechanical device},\ }\href@noop {} {\bibfield  {journal} {\bibinfo
  {journal} {Phys.~Rev.~X}\ }\textbf {\bibinfo {volume} {6}},\ \bibinfo {pages}
  {041024} (\bibinfo {year} {2016})}\BibitemShut {NoStop}%
\bibitem [{\citenamefont {Ruesink}\ \emph {et~al.}(2016)\citenamefont
  {Ruesink}, \citenamefont {Miri}, \citenamefont {Al{\`u}},\ and\ \citenamefont
  {Verhagen}}]{Verhagen2016isol}%
  \BibitemOpen
  \bibfield  {author} {\bibinfo {author} {\bibfnamefont {F.}~\bibnamefont
  {Ruesink}}, \bibinfo {author} {\bibfnamefont {M.-A.}\ \bibnamefont {Miri}},
  \bibinfo {author} {\bibfnamefont {A.}~\bibnamefont {Al{\`u}}},\ and\ \bibinfo
  {author} {\bibfnamefont {E.}~\bibnamefont {Verhagen}},\ }\bibfield  {title}
  {\bibinfo {title} {Nonreciprocity and magnetic-free isolation based on
  optomechanical interactions},\ }\href@noop {} {\bibfield  {journal} {\bibinfo
   {journal} {Nat. Commun.}\ }\textbf {\bibinfo {volume} {7}},\ \bibinfo
  {pages} {13662} (\bibinfo {year} {2016})}\BibitemShut {NoStop}%
\bibitem [{\citenamefont {Shen}\ \emph {et~al.}(2016)\citenamefont {Shen},
  \citenamefont {Zhang}, \citenamefont {Chen}, \citenamefont {Zou},
  \citenamefont {Xiao}, \citenamefont {Zou}, \citenamefont {Sun}, \citenamefont
  {Guo},\ and\ \citenamefont {Dong}}]{Kiina2016NonRes}%
  \BibitemOpen
  \bibfield  {author} {\bibinfo {author} {\bibfnamefont {Z.}~\bibnamefont
  {Shen}}, \bibinfo {author} {\bibfnamefont {Y.-L.}\ \bibnamefont {Zhang}},
  \bibinfo {author} {\bibfnamefont {Y.}~\bibnamefont {Chen}}, \bibinfo {author}
  {\bibfnamefont {C.-L.}\ \bibnamefont {Zou}}, \bibinfo {author} {\bibfnamefont
  {Y.-F.}\ \bibnamefont {Xiao}}, \bibinfo {author} {\bibfnamefont {X.-B.}\
  \bibnamefont {Zou}}, \bibinfo {author} {\bibfnamefont {F.-W.}\ \bibnamefont
  {Sun}}, \bibinfo {author} {\bibfnamefont {G.-C.}\ \bibnamefont {Guo}},\ and\
  \bibinfo {author} {\bibfnamefont {C.-H.}\ \bibnamefont {Dong}},\ }\bibfield
  {title} {\bibinfo {title} {Experimental realization of optomechanically
  induced non-reciprocity},\ }\href@noop {} {\bibfield  {journal} {\bibinfo
  {journal} {Nat. Photonics}\ }\textbf {\bibinfo {volume} {10}},\ \bibinfo
  {pages} {657} (\bibinfo {year} {2016})}\BibitemShut {NoStop}%
\bibitem [{\citenamefont {Mercier~de L\'epinay}\ \emph
  {et~al.}(2020)\citenamefont {Mercier~de L\'epinay}, \citenamefont
  {Ockeloen-Korppi}, \citenamefont {Malz},\ and\ \citenamefont
  {Sillanp\"a\"a}}]{Floquet2020}%
  \BibitemOpen
  \bibfield  {author} {\bibinfo {author} {\bibfnamefont {L.}~\bibnamefont
  {Mercier~de L\'epinay}}, \bibinfo {author} {\bibfnamefont {C.~F.}\
  \bibnamefont {Ockeloen-Korppi}}, \bibinfo {author} {\bibfnamefont
  {D.}~\bibnamefont {Malz}},\ and\ \bibinfo {author} {\bibfnamefont {M.~A.}\
  \bibnamefont {Sillanp\"a\"a}},\ }\bibfield  {title} {\bibinfo {title}
  {Nonreciprocal transport based on cavity {Floquet} modes in optomechanics},\
  }\href@noop {} {\bibfield  {journal} {\bibinfo  {journal} {Phys. Rev. Lett.}\
  }\textbf {\bibinfo {volume} {125}},\ \bibinfo {pages} {023603} (\bibinfo
  {year} {2020})}\BibitemShut {NoStop}%
\bibitem [{\citenamefont {Metzger}\ and\ \citenamefont
  {Karrai}(2004)}]{Karrai2004}%
  \BibitemOpen
  \bibfield  {author} {\bibinfo {author} {\bibfnamefont {C.~H.}\ \bibnamefont
  {Metzger}}\ and\ \bibinfo {author} {\bibfnamefont {K.}~\bibnamefont
  {Karrai}},\ }\bibfield  {title} {\bibinfo {title} {{Cavity cooling of a
  microlever}},\ }\href@noop {} {\bibfield  {journal} {\bibinfo  {journal}
  {Nature}\ }\textbf {\bibinfo {volume} {432}},\ \bibinfo {pages} {1002}
  (\bibinfo {year} {2004})}\BibitemShut {NoStop}%
\bibitem [{\citenamefont {Gigan}\ \emph {et~al.}(2006)\citenamefont {Gigan},
  \citenamefont {B{\"o}hm}, \citenamefont {Paternostro}, \citenamefont
  {Blaser}, \citenamefont {Langer}, \citenamefont {Hertzberg}, \citenamefont
  {Schwab}, \citenamefont {B{\"a}uerle}, \citenamefont {Aspelmeyer},\ and\
  \citenamefont {Zeilinger}}]{Aspelmeyer2006cool}%
  \BibitemOpen
  \bibfield  {author} {\bibinfo {author} {\bibfnamefont {S.}~\bibnamefont
  {Gigan}}, \bibinfo {author} {\bibfnamefont {H.~R.}\ \bibnamefont {B{\"o}hm}},
  \bibinfo {author} {\bibfnamefont {M.}~\bibnamefont {Paternostro}}, \bibinfo
  {author} {\bibfnamefont {F.}~\bibnamefont {Blaser}}, \bibinfo {author}
  {\bibfnamefont {G.}~\bibnamefont {Langer}}, \bibinfo {author} {\bibfnamefont
  {J.~B.}\ \bibnamefont {Hertzberg}}, \bibinfo {author} {\bibfnamefont {K.~C.}\
  \bibnamefont {Schwab}}, \bibinfo {author} {\bibfnamefont {D.}~\bibnamefont
  {B{\"a}uerle}}, \bibinfo {author} {\bibfnamefont {M.}~\bibnamefont
  {Aspelmeyer}},\ and\ \bibinfo {author} {\bibfnamefont {A.}~\bibnamefont
  {Zeilinger}},\ }\bibfield  {title} {\bibinfo {title} {{Self-cooling of a
  micromirror by radiation pressure}},\ }\href@noop {} {\bibfield  {journal}
  {\bibinfo  {journal} {Nature}\ }\textbf {\bibinfo {volume} {444}},\ \bibinfo
  {pages} {67} (\bibinfo {year} {2006})}\BibitemShut {NoStop}%
\bibitem [{\citenamefont {Thompson}\ \emph {et~al.}(2008)\citenamefont
  {Thompson}, \citenamefont {Zwickl}, \citenamefont {Jayich}, \citenamefont
  {Marquardt}, \citenamefont {Girvin},\ and\ \citenamefont
  {Harris}}]{thompson2008strong}%
  \BibitemOpen
  \bibfield  {author} {\bibinfo {author} {\bibfnamefont {J.}~\bibnamefont
  {Thompson}}, \bibinfo {author} {\bibfnamefont {B.}~\bibnamefont {Zwickl}},
  \bibinfo {author} {\bibfnamefont {A.}~\bibnamefont {Jayich}}, \bibinfo
  {author} {\bibfnamefont {F.}~\bibnamefont {Marquardt}}, \bibinfo {author}
  {\bibfnamefont {S.}~\bibnamefont {Girvin}},\ and\ \bibinfo {author}
  {\bibfnamefont {J.}~\bibnamefont {Harris}},\ }\bibfield  {title} {\bibinfo
  {title} {Strong dispersive coupling of a high-finesse cavity to a
  micromechanical membrane},\ }\href@noop {} {\bibfield  {journal} {\bibinfo
  {journal} {Nature}\ }\textbf {\bibinfo {volume} {452}},\ \bibinfo {pages}
  {72} (\bibinfo {year} {2008})}\BibitemShut {NoStop}%
\bibitem [{\citenamefont {Rossi}\ \emph {et~al.}(2018)\citenamefont {Rossi},
  \citenamefont {Mason}, \citenamefont {Chen}, \citenamefont {Tsaturyan},\ and\
  \citenamefont {Schliesser}}]{Schliesser2018FB}%
  \BibitemOpen
  \bibfield  {author} {\bibinfo {author} {\bibfnamefont {M.}~\bibnamefont
  {Rossi}}, \bibinfo {author} {\bibfnamefont {D.}~\bibnamefont {Mason}},
  \bibinfo {author} {\bibfnamefont {J.}~\bibnamefont {Chen}}, \bibinfo {author}
  {\bibfnamefont {Y.}~\bibnamefont {Tsaturyan}},\ and\ \bibinfo {author}
  {\bibfnamefont {A.}~\bibnamefont {Schliesser}},\ }\bibfield  {title}
  {\bibinfo {title} {Measurement-based quantum control of mechanical motion},\
  }\href@noop {} {\bibfield  {journal} {\bibinfo  {journal} {Nature}\ }\textbf
  {\bibinfo {volume} {563}},\ \bibinfo {pages} {53} (\bibinfo {year}
  {2018})}\BibitemShut {NoStop}%
\bibitem [{\citenamefont {Purdy}\ \emph {et~al.}(2010)\citenamefont {Purdy},
  \citenamefont {Brooks}, \citenamefont {Botter}, \citenamefont {Brahms},
  \citenamefont {Ma},\ and\ \citenamefont {Stamper-Kurn}}]{Stamper2010PRL}%
  \BibitemOpen
  \bibfield  {author} {\bibinfo {author} {\bibfnamefont {T.~P.}\ \bibnamefont
  {Purdy}}, \bibinfo {author} {\bibfnamefont {D.~W.~C.}\ \bibnamefont
  {Brooks}}, \bibinfo {author} {\bibfnamefont {T.}~\bibnamefont {Botter}},
  \bibinfo {author} {\bibfnamefont {N.}~\bibnamefont {Brahms}}, \bibinfo
  {author} {\bibfnamefont {Z.-Y.}\ \bibnamefont {Ma}},\ and\ \bibinfo {author}
  {\bibfnamefont {D.~M.}\ \bibnamefont {Stamper-Kurn}},\ }\bibfield  {title}
  {\bibinfo {title} {Tunable cavity optomechanics with ultracold atoms},\
  }\href@noop {} {\bibfield  {journal} {\bibinfo  {journal} {Phys.~Rev.~Lett.}\
  }\textbf {\bibinfo {volume} {105}},\ \bibinfo {pages} {133602} (\bibinfo
  {year} {2010})}\BibitemShut {NoStop}%
\bibitem [{\citenamefont {Regal}\ \emph {et~al.}(2008)\citenamefont {Regal},
  \citenamefont {Teufel},\ and\ \citenamefont {Lehnert}}]{Lehnert2008Nph}%
  \BibitemOpen
  \bibfield  {author} {\bibinfo {author} {\bibfnamefont {C.~A.}\ \bibnamefont
  {Regal}}, \bibinfo {author} {\bibfnamefont {J.~D.}\ \bibnamefont {Teufel}},\
  and\ \bibinfo {author} {\bibfnamefont {K.~W.}\ \bibnamefont {Lehnert}},\
  }\bibfield  {title} {\bibinfo {title} {{Measuring nanomechanical motion with
  a microwave cavity interferometer}},\ }\href@noop {} {\bibfield  {journal}
  {\bibinfo  {journal} {Nat. Phys.}\ }\textbf {\bibinfo {volume} {4}},\
  \bibinfo {pages} {555} (\bibinfo {year} {2008})}\BibitemShut {NoStop}%
\bibitem [{\citenamefont {Rocheleau}\ \emph {et~al.}(2010)\citenamefont
  {Rocheleau}, \citenamefont {Ndukum}, \citenamefont {Macklin}, \citenamefont
  {Hertzberg}, \citenamefont {Clerk},\ and\ \citenamefont
  {Schwab}}]{Schwab2010}%
  \BibitemOpen
  \bibfield  {author} {\bibinfo {author} {\bibfnamefont {T.}~\bibnamefont
  {Rocheleau}}, \bibinfo {author} {\bibfnamefont {T.}~\bibnamefont {Ndukum}},
  \bibinfo {author} {\bibfnamefont {C.}~\bibnamefont {Macklin}}, \bibinfo
  {author} {\bibfnamefont {J.~B.}\ \bibnamefont {Hertzberg}}, \bibinfo {author}
  {\bibfnamefont {A.~A.}\ \bibnamefont {Clerk}},\ and\ \bibinfo {author}
  {\bibfnamefont {K.~C.}\ \bibnamefont {Schwab}},\ }\bibfield  {title}
  {\bibinfo {title} {{Preparation and detection of a mechanical resonator near
  the ground state of motion}},\ }\href@noop {} {\bibfield  {journal} {\bibinfo
   {journal} {Nature}\ }\textbf {\bibinfo {volume} {463}},\ \bibinfo {pages}
  {72} (\bibinfo {year} {2010})}\BibitemShut {NoStop}%
\bibitem [{\citenamefont {Zhang}\ \emph {et~al.}(2016)\citenamefont {Zhang},
  \citenamefont {Zou}, \citenamefont {Jiang},\ and\ \citenamefont
  {Tang}}]{zhang2016cavity}%
  \BibitemOpen
  \bibfield  {author} {\bibinfo {author} {\bibfnamefont {X.}~\bibnamefont
  {Zhang}}, \bibinfo {author} {\bibfnamefont {C.-L.}\ \bibnamefont {Zou}},
  \bibinfo {author} {\bibfnamefont {L.}~\bibnamefont {Jiang}},\ and\ \bibinfo
  {author} {\bibfnamefont {H.~X.}\ \bibnamefont {Tang}},\ }\bibfield  {title}
  {\bibinfo {title} {Cavity magnomechanics},\ }\href@noop {} {\bibfield
  {journal} {\bibinfo  {journal} {Sci. Adv.}\ }\textbf {\bibinfo {volume}
  {2}},\ \bibinfo {pages} {e1501286} (\bibinfo {year} {2016})}\BibitemShut
  {NoStop}%
\bibitem [{\citenamefont {Gonzalez-Ballestero}\ \emph
  {et~al.}(2020{\natexlab{a}})\citenamefont {Gonzalez-Ballestero},
  \citenamefont {H{\"u}mmer}, \citenamefont {Gieseler},\ and\ \citenamefont
  {Romero-Isart}}]{gonzalez2020theory}%
  \BibitemOpen
  \bibfield  {author} {\bibinfo {author} {\bibfnamefont {C.}~\bibnamefont
  {Gonzalez-Ballestero}}, \bibinfo {author} {\bibfnamefont {D.}~\bibnamefont
  {H{\"u}mmer}}, \bibinfo {author} {\bibfnamefont {J.}~\bibnamefont
  {Gieseler}},\ and\ \bibinfo {author} {\bibfnamefont {O.}~\bibnamefont
  {Romero-Isart}},\ }\bibfield  {title} {\bibinfo {title} {Theory of quantum
  acoustomagnonics and acoustomechanics with a micromagnet},\ }\href@noop {}
  {\bibfield  {journal} {\bibinfo  {journal} {Phys. Rev. B}\ }\textbf {\bibinfo
  {volume} {101}},\ \bibinfo {pages} {125404} (\bibinfo {year}
  {2020}{\natexlab{a}})}\BibitemShut {NoStop}%
\bibitem [{\citenamefont {Gonzalez-Ballestero}\ \emph
  {et~al.}(2020{\natexlab{b}})\citenamefont {Gonzalez-Ballestero},
  \citenamefont {Gieseler},\ and\ \citenamefont
  {Romero-Isart}}]{gonzalez2020quantum}%
  \BibitemOpen
  \bibfield  {author} {\bibinfo {author} {\bibfnamefont {C.}~\bibnamefont
  {Gonzalez-Ballestero}}, \bibinfo {author} {\bibfnamefont {J.}~\bibnamefont
  {Gieseler}},\ and\ \bibinfo {author} {\bibfnamefont {O.}~\bibnamefont
  {Romero-Isart}},\ }\bibfield  {title} {\bibinfo {title} {Quantum
  acoustomechanics with a micromagnet},\ }\href@noop {} {\bibfield  {journal}
  {\bibinfo  {journal} {Phys. Rev. Lett.}\ }\textbf {\bibinfo {volume} {124}},\
  \bibinfo {pages} {093602} (\bibinfo {year} {2020}{\natexlab{b}})}\BibitemShut
  {NoStop}%
\bibitem [{\citenamefont {Potts}\ \emph {et~al.}(2021)\citenamefont {Potts},
  \citenamefont {Varga}, \citenamefont {Bittencourt}, \citenamefont
  {Kusminskiy},\ and\ \citenamefont {Davis}}]{potts2021dynamical}%
  \BibitemOpen
  \bibfield  {author} {\bibinfo {author} {\bibfnamefont {C.~A.}\ \bibnamefont
  {Potts}}, \bibinfo {author} {\bibfnamefont {E.}~\bibnamefont {Varga}},
  \bibinfo {author} {\bibfnamefont {V.~A. S.~V.}\ \bibnamefont {Bittencourt}},
  \bibinfo {author} {\bibfnamefont {S.~V.}\ \bibnamefont {Kusminskiy}},\ and\
  \bibinfo {author} {\bibfnamefont {J.~P.}\ \bibnamefont {Davis}},\ }\bibfield
  {title} {\bibinfo {title} {Dynamical backaction magnomechanics},\ }\href
  {https://doi.org/10.1103/PhysRevX.11.031053} {\bibfield  {journal} {\bibinfo
  {journal} {Phys. Rev. X}\ }\textbf {\bibinfo {volume} {11}},\ \bibinfo
  {pages} {031053} (\bibinfo {year} {2021})}\BibitemShut {NoStop}%
\bibitem [{\citenamefont {Kittel}(1948)}]{kittel1948fmr}%
  \BibitemOpen
  \bibfield  {author} {\bibinfo {author} {\bibfnamefont {C.}~\bibnamefont
  {Kittel}},\ }\bibfield  {title} {\bibinfo {title} {On the theory of
  ferromagnetic resonance absorption},\ }\href
  {https://doi.org/10.1103/PhysRev.73.155} {\bibfield  {journal} {\bibinfo
  {journal} {Phys. Rev.}\ }\textbf {\bibinfo {volume} {73}},\ \bibinfo {pages}
  {155} (\bibinfo {year} {1948})}\BibitemShut {NoStop}%
\bibitem [{\citenamefont {Kittel}(1949)}]{kittel1949physical}%
  \BibitemOpen
  \bibfield  {author} {\bibinfo {author} {\bibfnamefont {C.}~\bibnamefont
  {Kittel}},\ }\bibfield  {title} {\bibinfo {title} {Physical theory of
  ferromagnetic domains},\ }\href {https://doi.org/10.1103/RevModPhys.21.541}
  {\bibfield  {journal} {\bibinfo  {journal} {Rev. Mod. Phys.}\ }\textbf
  {\bibinfo {volume} {21}},\ \bibinfo {pages} {541} (\bibinfo {year}
  {1949})}\BibitemShut {NoStop}%
\bibitem [{\citenamefont {Heikkil{\"a}}(2013)}]{heikkila2013physics}%
  \BibitemOpen
  \bibfield  {author} {\bibinfo {author} {\bibfnamefont {T.~T.}\ \bibnamefont
  {Heikkil{\"a}}},\ }\href@noop {} {\emph {\bibinfo {title} {The physics of
  nanoelectronics: transport and fluctuation phenomena at low temperatures}}}\
  (\bibinfo  {publisher} {Oxford University Press},\ \bibinfo {year}
  {2013})\BibitemShut {NoStop}%
\bibitem [{\citenamefont {Landau}\ \emph {et~al.}(1986)\citenamefont {Landau},
  \citenamefont {Lifshitz}, \citenamefont {Kosevich}, \citenamefont {Sykes},
  \citenamefont {Pitaevskii},\ and\ \citenamefont {Reid}}]{landau1986theory}%
  \BibitemOpen
  \bibfield  {author} {\bibinfo {author} {\bibfnamefont {L.}~\bibnamefont
  {Landau}}, \bibinfo {author} {\bibfnamefont {E.}~\bibnamefont {Lifshitz}},
  \bibinfo {author} {\bibfnamefont {A.}~\bibnamefont {Kosevich}}, \bibinfo
  {author} {\bibfnamefont {J.}~\bibnamefont {Sykes}}, \bibinfo {author}
  {\bibfnamefont {L.}~\bibnamefont {Pitaevskii}},\ and\ \bibinfo {author}
  {\bibfnamefont {W.}~\bibnamefont {Reid}},\ }\href
  {https://books.google.fi/books?id=tpY-VkwCkAIC} {\emph {\bibinfo {title}
  {Theory of Elasticity: Volume 7}}},\ Course of theoretical physics\ (\bibinfo
   {publisher} {Elsevier Science},\ \bibinfo {year} {1986})\BibitemShut
  {NoStop}%
\bibitem [{\citenamefont {Reddy}\ and\ \citenamefont
  {Mahaffey}(2013)}]{reddy2013generalized}%
  \BibitemOpen
  \bibfield  {author} {\bibinfo {author} {\bibfnamefont {J.}~\bibnamefont
  {Reddy}}\ and\ \bibinfo {author} {\bibfnamefont {P.}~\bibnamefont
  {Mahaffey}},\ }\bibfield  {title} {\bibinfo {title} {Generalized beam
  theories accounting for von k{\'a}rm{\'a}n nonlinear strains with application
  to buckling},\ }\href@noop {} {\bibfield  {journal} {\bibinfo  {journal} {J.
  Coupled Syst. Multiscale Dyn.}\ }\textbf {\bibinfo {volume} {1}},\ \bibinfo
  {pages} {120} (\bibinfo {year} {2013})}\BibitemShut {NoStop}%
\bibitem [{\citenamefont {Nayfeh}\ and\ \citenamefont
  {Emam}(2008)}]{nayfeh2008exact}%
  \BibitemOpen
  \bibfield  {author} {\bibinfo {author} {\bibfnamefont {A.~H.}\ \bibnamefont
  {Nayfeh}}\ and\ \bibinfo {author} {\bibfnamefont {S.~A.}\ \bibnamefont
  {Emam}},\ }\bibfield  {title} {\bibinfo {title} {Exact solution and stability
  of postbuckling configurations of beams},\ }\href@noop {} {\bibfield
  {journal} {\bibinfo  {journal} {Nonlinear Dyn.}\ }\textbf {\bibinfo {volume}
  {54}},\ \bibinfo {pages} {395} (\bibinfo {year} {2008})}\BibitemShut
  {NoStop}%
\bibitem [{num()}]{numericsnote}%
  \BibitemOpen
  \href@noop {} {}\bibinfo {note} {All numerics was constructed with Python
  and, more precisely, with numpy-, scipy-, and matplotlib-packages. The
  scripts to produce Figs.~\ref{fig:beta1} -- \ref{fig:thetavsphi} are
  available at
  \url{https://gitlab.jyu.fi/jyucmt/suspended-beam-magnomechanics}.}\BibitemShut
  {Stop}%
\bibitem [{\citenamefont {Stoner}\ and\ \citenamefont
  {Wohlfarth}(1948)}]{stoner1948mechanism}%
  \BibitemOpen
  \bibfield  {author} {\bibinfo {author} {\bibfnamefont {E.~C.}\ \bibnamefont
  {Stoner}}\ and\ \bibinfo {author} {\bibfnamefont {E.}~\bibnamefont
  {Wohlfarth}},\ }\bibfield  {title} {\bibinfo {title} {A mechanism of magnetic
  hysteresis in heterogeneous alloys},\ }\href@noop {} {\bibfield  {journal}
  {\bibinfo  {journal} {Philos. Trans. R. Soc. A}\ }\textbf {\bibinfo {volume}
  {240}},\ \bibinfo {pages} {599} (\bibinfo {year} {1948})}\BibitemShut
  {NoStop}%
\bibitem [{\citenamefont {Gilbert}(2004)}]{gilbert2004phenomenological}%
  \BibitemOpen
  \bibfield  {author} {\bibinfo {author} {\bibfnamefont {T.~L.}\ \bibnamefont
  {Gilbert}},\ }\bibfield  {title} {\bibinfo {title} {A phenomenological theory
  of damping in ferromagnetic materials},\ }\href@noop {} {\bibfield  {journal}
  {\bibinfo  {journal} {IEEE Trans. Magn.}\ }\textbf {\bibinfo {volume} {40}},\
  \bibinfo {pages} {3443} (\bibinfo {year} {2004})}\BibitemShut {NoStop}%
\bibitem [{can()}]{cantilevernote}%
  \BibitemOpen
  \href@noop {} {}\bibinfo {note} {In a cantilever, the free energy term $\mu_0
  M_x M_z u'(z)$ due to demagnetizing
  field~\eqref{eq:demagnetizing_field_energy} does not disappear. Rather,
  averaging over the length gives an additional free energy term $-M_x M_z
  [u(L) - u(0)]/L$ to Eq.~\eqref{eq:megn_energy_eff_field} which may change the
  form of the magnomechanical coupling.}\BibitemShut {Stop}%
\bibitem [{\citenamefont {Sander}(1999)}]{sander1999correlation}%
  \BibitemOpen
  \bibfield  {author} {\bibinfo {author} {\bibfnamefont {D.}~\bibnamefont
  {Sander}},\ }\bibfield  {title} {\bibinfo {title} {The correlation between
  mechanical stress and magnetic anisotropy in ultrathin films},\ }\href@noop
  {} {\bibfield  {journal} {\bibinfo  {journal} {Rep. Prog. Phys.}\ }\textbf
  {\bibinfo {volume} {62}},\ \bibinfo {pages} {809} (\bibinfo {year}
  {1999})}\BibitemShut {NoStop}%
\bibitem [{str()}]{strainnote}%
  \BibitemOpen
  \href@noop {} {}\bibinfo {note} {The strain tensor $\epsilon_{ij}$ connects
  the lengths of line elements before and after
  displacement~\cite{landau1986theory}: Let $dL = \sqrt{dx_1^2 + dx_2^2 +
  dx_3^2}$ be the infinitesimal distance between two points on the beam at
  $\bm{r}$. Then, after displacing the beam by a vector field $\bm{u}$, the
  points are at $\bm{r'} = \bm{r} + \bm{u}(\bm{r})$ and the lengths of the line
  elements are changed to $dx'_i = dx_i + (\nabla{\bm{u}}\cdot \bm{dx})_i$.
  Consequently, $dL'^2 = dL^2 + 2 \sum_{ij}\epsilon_{ij}dx_i dx_j$ with
  $\epsilon_{ij}$ as defined in the main text.}\BibitemShut {Stop}%
\bibitem [{\citenamefont {Serafini}(2017)}]{serafini2017quantum}%
  \BibitemOpen
  \bibfield  {author} {\bibinfo {author} {\bibfnamefont {A.}~\bibnamefont
  {Serafini}},\ }\href {https://books.google.fi/books?id=zHtgvgAACAAJ} {\emph
  {\bibinfo {title} {Quantum Continuous Variables: A Primer of Theoretical
  Methods}}}\ (\bibinfo  {publisher} {CRC Press, Taylor \& Francis Group},\
  \bibinfo {year} {2017})\BibitemShut {NoStop}%
\bibitem [{\citenamefont {Zou}\ \emph {et~al.}(2020)\citenamefont {Zou},
  \citenamefont {Kim},\ and\ \citenamefont {Tserkovnyak}}]{zou2020tuning}%
  \BibitemOpen
  \bibfield  {author} {\bibinfo {author} {\bibfnamefont {J.}~\bibnamefont
  {Zou}}, \bibinfo {author} {\bibfnamefont {S.~K.}\ \bibnamefont {Kim}},\ and\
  \bibinfo {author} {\bibfnamefont {Y.}~\bibnamefont {Tserkovnyak}},\
  }\bibfield  {title} {\bibinfo {title} {Tuning entanglement by squeezing
  magnons in anisotropic magnets},\ }\href
  {https://doi.org/10.1103/PhysRevB.101.014416} {\bibfield  {journal} {\bibinfo
   {journal} {Phys. Rev. B}\ }\textbf {\bibinfo {volume} {101}},\ \bibinfo
  {pages} {014416} (\bibinfo {year} {2020})}\BibitemShut {NoStop}%
\bibitem [{\citenamefont {Sharma}\ \emph {et~al.}(2021)\citenamefont {Sharma},
  \citenamefont {Bittencourt}, \citenamefont {Karenowska},\ and\ \citenamefont
  {Kusminskiy}}]{sharma2021spin}%
  \BibitemOpen
  \bibfield  {author} {\bibinfo {author} {\bibfnamefont {S.}~\bibnamefont
  {Sharma}}, \bibinfo {author} {\bibfnamefont {V.~A.}\ \bibnamefont
  {Bittencourt}}, \bibinfo {author} {\bibfnamefont {A.~D.}\ \bibnamefont
  {Karenowska}},\ and\ \bibinfo {author} {\bibfnamefont {S.~V.}\ \bibnamefont
  {Kusminskiy}},\ }\bibfield  {title} {\bibinfo {title} {Spin cat states in
  ferromagnetic insulators},\ }\href@noop {} {\bibfield  {journal} {\bibinfo
  {journal} {Physical Review B}\ }\textbf {\bibinfo {volume} {103}},\ \bibinfo
  {pages} {L100403} (\bibinfo {year} {2021})}\BibitemShut {NoStop}%
\bibitem [{\citenamefont {Kamra}\ and\ \citenamefont
  {Belzig}(2016)}]{kamra2016super}%
  \BibitemOpen
  \bibfield  {author} {\bibinfo {author} {\bibfnamefont {A.}~\bibnamefont
  {Kamra}}\ and\ \bibinfo {author} {\bibfnamefont {W.}~\bibnamefont {Belzig}},\
  }\bibfield  {title} {\bibinfo {title} {Super-poissonian shot noise of
  squeezed-magnon mediated spin transport},\ }\href
  {https://doi.org/10.1103/PhysRevLett.116.146601} {\bibfield  {journal}
  {\bibinfo  {journal} {Phys. Rev. Lett.}\ }\textbf {\bibinfo {volume} {116}},\
  \bibinfo {pages} {146601} (\bibinfo {year} {2016})}\BibitemShut {NoStop}%
\bibitem [{\citenamefont {Gardiner}\ and\ \citenamefont
  {Collett}(1985)}]{gardiner1985input}%
  \BibitemOpen
  \bibfield  {author} {\bibinfo {author} {\bibfnamefont {C.~W.}\ \bibnamefont
  {Gardiner}}\ and\ \bibinfo {author} {\bibfnamefont {M.~J.}\ \bibnamefont
  {Collett}},\ }\bibfield  {title} {\bibinfo {title} {Input and output in
  damped quantum systems: Quantum stochastic differential equations and the
  master equation},\ }\href@noop {} {\bibfield  {journal} {\bibinfo  {journal}
  {Phys. Rev. A}\ }\textbf {\bibinfo {volume} {31}},\ \bibinfo {pages} {3761}
  (\bibinfo {year} {1985})}\BibitemShut {NoStop}%
\bibitem [{\citenamefont {McKenzie-Sell}\ \emph {et~al.}(2019)\citenamefont
  {McKenzie-Sell}, \citenamefont {Xie}, \citenamefont {Lee}, \citenamefont
  {Robinson}, \citenamefont {Ciccarelli},\ and\ \citenamefont
  {Haigh}}]{mckenziesell2019low}%
  \BibitemOpen
  \bibfield  {author} {\bibinfo {author} {\bibfnamefont {L.}~\bibnamefont
  {McKenzie-Sell}}, \bibinfo {author} {\bibfnamefont {J.}~\bibnamefont {Xie}},
  \bibinfo {author} {\bibfnamefont {C.-M.}\ \bibnamefont {Lee}}, \bibinfo
  {author} {\bibfnamefont {J.~W.~A.}\ \bibnamefont {Robinson}}, \bibinfo
  {author} {\bibfnamefont {C.}~\bibnamefont {Ciccarelli}},\ and\ \bibinfo
  {author} {\bibfnamefont {J.~A.}\ \bibnamefont {Haigh}},\ }\bibfield  {title}
  {\bibinfo {title} {Low-impedance superconducting microwave resonators for
  strong coupling to small magnetic mode volumes},\ }\href
  {https://doi.org/10.1103/PhysRevB.99.140414} {\bibfield  {journal} {\bibinfo
  {journal} {Phys. Rev. B}\ }\textbf {\bibinfo {volume} {99}},\ \bibinfo
  {pages} {140414} (\bibinfo {year} {2019})}\BibitemShut {NoStop}%
\bibitem [{\citenamefont {Cleland}\ and\ \citenamefont
  {Roukes}(1996)}]{cleland1996fabrication}%
  \BibitemOpen
  \bibfield  {author} {\bibinfo {author} {\bibfnamefont {A.~N.}\ \bibnamefont
  {Cleland}}\ and\ \bibinfo {author} {\bibfnamefont {M.~L.}\ \bibnamefont
  {Roukes}},\ }\bibfield  {title} {\bibinfo {title} {{Fabrication of high
  frequency nanometer scale mechanical resonators from bulk Si crystals}},\
  }\href@noop {} {\bibfield  {journal} {\bibinfo  {journal} {Appl. Phys.
  Lett.}\ }\textbf {\bibinfo {volume} {69}},\ \bibinfo {pages} {2653} (\bibinfo
  {year} {1996})}\BibitemShut {NoStop}%
\bibitem [{\citenamefont {Greywall}\ \emph {et~al.}(1996)\citenamefont
  {Greywall}, \citenamefont {Yurke}, \citenamefont {Busch},\ and\ \citenamefont
  {Arney}}]{Yurke1996Magnetomot}%
  \BibitemOpen
  \bibfield  {author} {\bibinfo {author} {\bibfnamefont {D.~S.}\ \bibnamefont
  {Greywall}}, \bibinfo {author} {\bibfnamefont {B.}~\bibnamefont {Yurke}},
  \bibinfo {author} {\bibfnamefont {P.~A.}\ \bibnamefont {Busch}},\ and\
  \bibinfo {author} {\bibfnamefont {S.~C.}\ \bibnamefont {Arney}},\ }\bibfield
  {title} {\bibinfo {title} {Low-temperature anomalies in the dissipation of
  small mechanical resonators},\ }\href@noop {} {\bibfield  {journal} {\bibinfo
   {journal} {Europhys. Lett.}\ }\textbf {\bibinfo {volume} {34}},\ \bibinfo
  {pages} {37} (\bibinfo {year} {1996})}\BibitemShut {NoStop}%
\bibitem [{\citenamefont {Li}\ \emph {et~al.}(2008)\citenamefont {Li},
  \citenamefont {Pashkin}, \citenamefont {Astafiev}, \citenamefont {Nakamura},
  \citenamefont {Tsai},\ and\ \citenamefont {Im}}]{Pashkin08}%
  \BibitemOpen
  \bibfield  {author} {\bibinfo {author} {\bibfnamefont {T.~F.}\ \bibnamefont
  {Li}}, \bibinfo {author} {\bibfnamefont {Y.~A.}\ \bibnamefont {Pashkin}},
  \bibinfo {author} {\bibfnamefont {O.}~\bibnamefont {Astafiev}}, \bibinfo
  {author} {\bibfnamefont {Y.}~\bibnamefont {Nakamura}}, \bibinfo {author}
  {\bibfnamefont {J.~S.}\ \bibnamefont {Tsai}},\ and\ \bibinfo {author}
  {\bibfnamefont {H.}~\bibnamefont {Im}},\ }\bibfield  {title} {\bibinfo
  {title} {High-frequency metallic nanomechanical resonators},\ }\href@noop {}
  {\bibfield  {journal} {\bibinfo  {journal} {Appl. Phys. Lett.}\ }\textbf
  {\bibinfo {volume} {92}},\ \bibinfo {pages} {043112} (\bibinfo {year}
  {2008})}\BibitemShut {NoStop}%
\bibitem [{\citenamefont {Marquardt}\ \emph {et~al.}(2007)\citenamefont
  {Marquardt}, \citenamefont {Chen}, \citenamefont {Clerk},\ and\ \citenamefont
  {Girvin}}]{marquardt2007quantum}%
  \BibitemOpen
  \bibfield  {author} {\bibinfo {author} {\bibfnamefont {F.}~\bibnamefont
  {Marquardt}}, \bibinfo {author} {\bibfnamefont {J.~P.}\ \bibnamefont {Chen}},
  \bibinfo {author} {\bibfnamefont {A.~A.}\ \bibnamefont {Clerk}},\ and\
  \bibinfo {author} {\bibfnamefont {S.~M.}\ \bibnamefont {Girvin}},\ }\bibfield
   {title} {\bibinfo {title} {Quantum theory of cavity-assisted sideband
  cooling of mechanical motion},\ }\href
  {https://doi.org/10.1103/PhysRevLett.99.093902} {\bibfield  {journal}
  {\bibinfo  {journal} {Phys. Rev. Lett.}\ }\textbf {\bibinfo {volume} {99}},\
  \bibinfo {pages} {093902} (\bibinfo {year} {2007})}\BibitemShut {NoStop}%
\bibitem [{\citenamefont {Massel}\ \emph {et~al.}(2011)\citenamefont {Massel},
  \citenamefont {Heikkil{\"a}}, \citenamefont {Pirkkalainen}, \citenamefont
  {Cho}, \citenamefont {Saloniemi}, \citenamefont {Hakonen},\ and\
  \citenamefont {Sillanp{\"a}{\"a}}}]{MechAmpPaper}%
  \BibitemOpen
  \bibfield  {author} {\bibinfo {author} {\bibfnamefont {F.}~\bibnamefont
  {Massel}}, \bibinfo {author} {\bibfnamefont {T.~T.}\ \bibnamefont
  {Heikkil{\"a}}}, \bibinfo {author} {\bibfnamefont {J.-M.}\ \bibnamefont
  {Pirkkalainen}}, \bibinfo {author} {\bibfnamefont {S.~U.}\ \bibnamefont
  {Cho}}, \bibinfo {author} {\bibfnamefont {H.}~\bibnamefont {Saloniemi}},
  \bibinfo {author} {\bibfnamefont {P.~J.}\ \bibnamefont {Hakonen}},\ and\
  \bibinfo {author} {\bibfnamefont {M.~A.}\ \bibnamefont {Sillanp{\"a}{\"a}}},\
  }\bibfield  {title} {\bibinfo {title} {Microwave amplification with
  nanomechanical resonators},\ }\href@noop {} {\bibfield  {journal} {\bibinfo
  {journal} {Nature}\ }\textbf {\bibinfo {volume} {480}},\ \bibinfo {pages}
  {351} (\bibinfo {year} {2011})}\BibitemShut {NoStop}%
\bibitem [{\citenamefont {Botter}\ \emph {et~al.}(2012)\citenamefont {Botter},
  \citenamefont {Brooks}, \citenamefont {Brahms}, \citenamefont {Schreppler},\
  and\ \citenamefont {Stamper-Kurn}}]{botter2012linear}%
  \BibitemOpen
  \bibfield  {author} {\bibinfo {author} {\bibfnamefont {T.}~\bibnamefont
  {Botter}}, \bibinfo {author} {\bibfnamefont {D.~W.~C.}\ \bibnamefont
  {Brooks}}, \bibinfo {author} {\bibfnamefont {N.}~\bibnamefont {Brahms}},
  \bibinfo {author} {\bibfnamefont {S.}~\bibnamefont {Schreppler}},\ and\
  \bibinfo {author} {\bibfnamefont {D.~M.}\ \bibnamefont {Stamper-Kurn}},\
  }\bibfield  {title} {\bibinfo {title} {Linear amplifier model for
  optomechanical systems},\ }\href {https://doi.org/10.1103/PhysRevA.85.013812}
  {\bibfield  {journal} {\bibinfo  {journal} {Phys. Rev. A}\ }\textbf {\bibinfo
  {volume} {85}},\ \bibinfo {pages} {013812} (\bibinfo {year}
  {2012})}\BibitemShut {NoStop}%
\bibitem [{\citenamefont {Cohen}\ \emph {et~al.}(2020)\citenamefont {Cohen},
  \citenamefont {Bothner}, \citenamefont {Blanter},\ and\ \citenamefont
  {Steele}}]{cohen2020optomechanical}%
  \BibitemOpen
  \bibfield  {author} {\bibinfo {author} {\bibfnamefont {M.~A.}\ \bibnamefont
  {Cohen}}, \bibinfo {author} {\bibfnamefont {D.}~\bibnamefont {Bothner}},
  \bibinfo {author} {\bibfnamefont {Y.~M.}\ \bibnamefont {Blanter}},\ and\
  \bibinfo {author} {\bibfnamefont {G.~A.}\ \bibnamefont {Steele}},\ }\bibfield
   {title} {\bibinfo {title} {Optomechanical microwave amplification without
  mechanical amplification},\ }\href
  {https://doi.org/10.1103/PhysRevApplied.13.014028} {\bibfield  {journal}
  {\bibinfo  {journal} {Phys. Rev. Appl.}\ }\textbf {\bibinfo {volume} {13}},\
  \bibinfo {pages} {014028} (\bibinfo {year} {2020})}\BibitemShut {NoStop}%
\bibitem [{\citenamefont {Losby}\ \emph {et~al.}(2010)\citenamefont {Losby},
  \citenamefont {Burgess}, \citenamefont {Holt}, \citenamefont {Westwood},
  \citenamefont {Mitlin}, \citenamefont {Hiebert},\ and\ \citenamefont
  {Freeman}}]{Freeman010MagCantil}%
  \BibitemOpen
  \bibfield  {author} {\bibinfo {author} {\bibfnamefont {J.}~\bibnamefont
  {Losby}}, \bibinfo {author} {\bibfnamefont {J.~A.~J.}\ \bibnamefont
  {Burgess}}, \bibinfo {author} {\bibfnamefont {C.~M.~B.}\ \bibnamefont
  {Holt}}, \bibinfo {author} {\bibfnamefont {J.~N.}\ \bibnamefont {Westwood}},
  \bibinfo {author} {\bibfnamefont {D.}~\bibnamefont {Mitlin}}, \bibinfo
  {author} {\bibfnamefont {W.~K.}\ \bibnamefont {Hiebert}},\ and\ \bibinfo
  {author} {\bibfnamefont {M.~R.}\ \bibnamefont {Freeman}},\ }\bibfield
  {title} {\bibinfo {title} {Nanomechanical torque magnetometry of permalloy
  cantilevers},\ }\href@noop {} {\bibfield  {journal} {\bibinfo  {journal} {J.
  Appl. Phys.}\ }\textbf {\bibinfo {volume} {108}},\ \bibinfo {pages} {123910}
  (\bibinfo {year} {2010})}\BibitemShut {NoStop}%
\bibitem [{\citenamefont {Arisawa}\ \emph {et~al.}(2019)\citenamefont
  {Arisawa}, \citenamefont {Daimon}, \citenamefont {Oikawa}, \citenamefont
  {Seo}, \citenamefont {Harii}, \citenamefont {Oyanagi},\ and\ \citenamefont
  {Saitoh}}]{Arisawa2019deltaE}%
  \BibitemOpen
  \bibfield  {author} {\bibinfo {author} {\bibfnamefont {H.}~\bibnamefont
  {Arisawa}}, \bibinfo {author} {\bibfnamefont {S.}~\bibnamefont {Daimon}},
  \bibinfo {author} {\bibfnamefont {Y.}~\bibnamefont {Oikawa}}, \bibinfo
  {author} {\bibfnamefont {Y.-J.}\ \bibnamefont {Seo}}, \bibinfo {author}
  {\bibfnamefont {K.}~\bibnamefont {Harii}}, \bibinfo {author} {\bibfnamefont
  {K.}~\bibnamefont {Oyanagi}},\ and\ \bibinfo {author} {\bibfnamefont
  {E.}~\bibnamefont {Saitoh}},\ }\bibfield  {title} {\bibinfo {title}
  {{Magnetomechanical sensing based on delta-E effect in Y$_3$Fe$_5$O$_{12}$
  micro bridge}},\ }\href@noop {} {\bibfield  {journal} {\bibinfo  {journal}
  {Appl. Phys. Lett.}\ }\textbf {\bibinfo {volume} {114}},\ \bibinfo {pages}
  {122402} (\bibinfo {year} {2019})}\BibitemShut {NoStop}%
\bibitem [{\citenamefont {Heyroth}\ \emph {et~al.}(2019)\citenamefont
  {Heyroth}, \citenamefont {Hauser}, \citenamefont {Trempler}, \citenamefont
  {Geyer}, \citenamefont {Syrowatka}, \citenamefont {Dreyer}, \citenamefont
  {Ebbinghaus}, \citenamefont {Woltersdorf},\ and\ \citenamefont
  {Schmidt}}]{Schmidt2019YIGbridge}%
  \BibitemOpen
  \bibfield  {author} {\bibinfo {author} {\bibfnamefont {F.}~\bibnamefont
  {Heyroth}}, \bibinfo {author} {\bibfnamefont {C.}~\bibnamefont {Hauser}},
  \bibinfo {author} {\bibfnamefont {P.}~\bibnamefont {Trempler}}, \bibinfo
  {author} {\bibfnamefont {P.}~\bibnamefont {Geyer}}, \bibinfo {author}
  {\bibfnamefont {F.}~\bibnamefont {Syrowatka}}, \bibinfo {author}
  {\bibfnamefont {R.}~\bibnamefont {Dreyer}}, \bibinfo {author} {\bibfnamefont
  {S.}~\bibnamefont {Ebbinghaus}}, \bibinfo {author} {\bibfnamefont
  {G.}~\bibnamefont {Woltersdorf}},\ and\ \bibinfo {author} {\bibfnamefont
  {G.}~\bibnamefont {Schmidt}},\ }\bibfield  {title} {\bibinfo {title}
  {Monocrystalline freestanding three-dimensional yttrium-iron-garnet magnon
  nanoresonators},\ }\href@noop {} {\bibfield  {journal} {\bibinfo  {journal}
  {Phys. Rev. Appl.}\ }\textbf {\bibinfo {volume} {12}},\ \bibinfo {pages}
  {054031} (\bibinfo {year} {2019})}\BibitemShut {NoStop}%
\bibitem [{\citenamefont {Gowtham}\ \emph {et~al.}(2016)\citenamefont
  {Gowtham}, \citenamefont {Stiehl}, \citenamefont {Ralph},\ and\ \citenamefont
  {Buhrman}}]{gowtham2016thickness}%
  \BibitemOpen
  \bibfield  {author} {\bibinfo {author} {\bibfnamefont {P.~G.}\ \bibnamefont
  {Gowtham}}, \bibinfo {author} {\bibfnamefont {G.~M.}\ \bibnamefont {Stiehl}},
  \bibinfo {author} {\bibfnamefont {D.~C.}\ \bibnamefont {Ralph}},\ and\
  \bibinfo {author} {\bibfnamefont {R.~A.}\ \bibnamefont {Buhrman}},\
  }\bibfield  {title} {\bibinfo {title} {Thickness-dependent magnetoelasticity
  and its effects on perpendicular magnetic anisotropy in ta/cofeb/mgo thin
  films},\ }\href {https://doi.org/10.1103/PhysRevB.93.024404} {\bibfield
  {journal} {\bibinfo  {journal} {Phys. Rev. B}\ }\textbf {\bibinfo {volume}
  {93}},\ \bibinfo {pages} {024404} (\bibinfo {year} {2016})}\BibitemShut
  {NoStop}%
\bibitem [{\citenamefont {Wohlfarth}(1958)}]{wohlfarth1958relations}%
  \BibitemOpen
  \bibfield  {author} {\bibinfo {author} {\bibfnamefont {E.~P.}\ \bibnamefont
  {Wohlfarth}},\ }\bibfield  {title} {\bibinfo {title} {Relations between
  different modes of acquisition of the remanent magnetization of ferromagnetic
  particles},\ }\href@noop {} {\bibfield  {journal} {\bibinfo  {journal} {J.
  Appl. Phys.}\ }\textbf {\bibinfo {volume} {29}},\ \bibinfo {pages} {595}
  (\bibinfo {year} {1958})}\BibitemShut {NoStop}%
\bibitem [{\citenamefont {Stancil}\ and\ \citenamefont
  {Prabhakar}(2009)}]{stancil2009spin}%
  \BibitemOpen
  \bibfield  {author} {\bibinfo {author} {\bibfnamefont {D.}~\bibnamefont
  {Stancil}}\ and\ \bibinfo {author} {\bibfnamefont {A.}~\bibnamefont
  {Prabhakar}},\ }\href {https://books.google.fi/books?id=ehN6-ubvKwoC} {\emph
  {\bibinfo {title} {Spin Waves: Theory and Applications}}}\ (\bibinfo
  {publisher} {Springer US},\ \bibinfo {year} {2009})\BibitemShut {NoStop}%
\end{thebibliography}
\end{document}